\documentclass[fleqn,usenatbib]{mnras}

\usepackage{newtxtext,newtxmath}

\usepackage[T1]{fontenc}

\DeclareRobustCommand{\VAN}[3]{#2}
\let\VANthebibliography\thebibliography
\def\thebibliography{\DeclareRobustCommand{\VAN}[3]{##3}\VANthebibliography}

\usepackage{graphicx}	
\usepackage{amsmath}	
\usepackage{adjustbox}

\title[AGN classification comparison]{The Close AGN Reference Survey (CARS):
A comparison between sub-mm and optical AGN diagnostic diagrams}

\author[J. S. Elford et al]{\parbox{\textwidth}{
Jacob S. Elford,$^{1,2}$\thanks{E-mail: jacob.elford@mail.udp.cl}
Timothy A. Davis,$^{1}$
Ilaria Ruffa,$^{1,3}$
Stefi A. Baum,$^{4,5}$
Francoise Combes,$^{6}$
Massimo Gaspari,$^{7}$
Rebecca McElroy,$^{8}$
Christopher P. O'Dea,$^{4,5}$
Osase Omoruyi,$^{9}$
Mainak Singha,$^{10,11,12}$
Grant R.~Tremblay,$^{9}$
Nico Winkel$^{13}$}\vspace{0.3cm}
\\
$^{1}$Cardiff Hub for Astrophysics Research \&\ Technology, School of Physics \&\ Astronomy, Cardiff University, Queens Buildings, The Parade, Cardiff, CF24 3AA, UK\\
$^{2}$Instituto de Estudios Astrof\'{i}sicos, Facultad de Ingenier\'{i}a y Ciencias, Universidad Diego Portales, Av. Ej\'{e}rcito Libertador 441, Santiago, Chile\\
$^{3}$INAF, Arcetri Astrophysical Observatory, Largo Enrico Fermi 5, I-50125 Florence, Italy\\
$^{4}$Center for Space Plasma \& Aeronomic Research 
University of Alabama in Huntsville
Huntsville, AL 35899, USA\\
$^{5}$Department of Physics \& Astronomy, University of Manitoba, 30A Sifton Road, Winnipeg, MB R3T 2N2, Canada\\
$^{6}$LUX, Observatoire de Paris, PSL Univ., Coll\`{e}ge de France, CNRS, Sorbonne Univ., Paris, France\\
$^{7}$Department of Physics, Informatics \& Mathematics, University of Modena \& Reggio Emilia, 41125 MO, IT\\
$^{8}$Centre for Astrophysics, University of Southern Queensland, Toowoomba, QLD 4350, Australia\\
$^{9}$Center for Astrophysics $|$ Harvard \& Smithsonian, 60 Garden St., Cambridge, MA 02138, USA\\
$^{10}$Astrophysics Science Division, NASA, Goddard Space Flight Center, Greenbelt, MD 20771, USA\\
$^{11}$Department of Physics, The Catholic University of America, Washington, DC 20064, USA\\
$^{12}$Center for Research and Exploration in Space Science and Technology, NASA Goddard Space Flight Center, Greenbelt, MD 20771, USA\\
$^{13}$Max-Planck-Institut f\"{u}r Astronomie, K\"{o}nigstuhl 17, D-69117 Heidelberg, Germany\\
% ...
}

\date{Accepted XXX. Received YYY; in original form ZZZ}

\pubyear{2025}

\begin{document}
\label{firstpage}
\pagerange{\pageref{firstpage}--\pageref{lastpage}}
\maketitle

\begin{abstract}
The $L_{\rm IR}-L_{\rm HCN}$ relation suggests that there is a tight connection between dense gas and star formation. We use data from the Close AGN Reference Survey (CARS) to investigate the dense gas - star formation relation in AGN hosting galaxies, and the use of dense gas as an active galactic nuclei (AGN) diagnostic. Our sample contains five Type-1 (unobscured) AGN that were observed with the Atacama Large Millimeter/submillimeter Array (ALMA) with the aim to detect HCN(4-3), HCO$^+$(4-3) and CS(7-6).  We detect the dense gas emission required for this analysis in 3 of the 5 targets. We find that despite the potential impact from the AGN on the line fluxes of these sources, they still follow the $L_{\rm IR}-L_{\rm HCN}$ relation. We then go on to test claims that the HCN/HCO+ and HCN/CS line ratios can be used as a tool to classify AGN in the \textit{sub-mm HCN diagram}. We produce the classic ionised emission-line ratio diagnostics (the so-called BPT diagrams), using available CARS data from the Multi Unit Spectroscopic Explorer (MUSE). We then compare the BPT classification with the sub-mm classification made using the dense gas tracers. Where it was possible to complete the analysis we find general agreement between optical and sub-mm classified gas excitation mechanisms. This suggests that AGN can contribute to the excitation of both the low density gas in the warm ionised medium and the high density gas in molecular clouds simultaneously, perhaps through X-ray, cosmic ray or shock heating mechanisms. 

\end{abstract}  
\begin{keywords}
galaxies: active -- galaxies: ISM -- galaxies: nuclei
\end{keywords}



\section{Introduction}
Supermassive black holes (SMBHs) are found in the centre of nearly all galaxies with stellar masses $M_{*}\gtrsim10^{9}\,\rm M_{\odot}$. 
Tight correlations between the SMBH mass and host galaxy properties -- such as bulge mass \citep[e.g.][]{Magorrian1998,MarconiHunt2004}, stellar velocity dispersion \citep[e.g.][]{FerrareseMerritt2000,Tremaine2002,Gultekin2009}, and X-ray halo temperature/luminosity \citep{Gaspari19} -- suggest a self-regulated co-evolution between SMBHs and galaxies \citep[e.g.][]{KormendyHo2013}. 
Accreting SMBHs, also known as Active Galactic Nuclei (AGN), release substantial amount of energy back to the interstellar/circumgalactic medium (ISM/CGM), are thought to significantly influence their host galaxies evolution. The multiphase ISM/CGM is believed to play a critical role in feeding the AGN \citep[e.g.][]{Shlosman1990,Gasparia13,Tremblay16,Izumi16,Temi18,Maccagni21,Ruffa22,McKinley22,Maccagni23}, but the extent to which AGN feedback impacts the environment via shocks, turbulence, and uplift over multiple scales remains a subject of active research \citep[e.g.][]{Bower2006,Croton2006,KingPounds2015,Morganti2017,Gaspari20,Harrison24}. 
\par
Thanks to the latest-generation (sub-)millimetre interferometers such as the Atacama Large Millimeter/submillimeter Array (ALMA), the cold component of the ISM in galaxies has been now probed using a variety of different molecular gas tracers, including CO, HCN, $\rm HCO^+$ and CS. CO is the workhorse for molecular gas studies in galaxies, as it is the most abundant molecule after H$_{2}$ and has excitation temperatures that make it easily excited even at low temperatures (i.e.\,$T_{\rm ex}=5.53$~K for the ground $J =1\rightarrow0$ transition; \citealp[e.g.][]{Bolatto13}). A number of studies have detected different CO transitions in the nuclear regions of AGN-host galaxies \citep[e.g.][]{Santi2014,Moser2016,Oosterloo17,Ruffa19,Koss2021,Ruffa22,Lelli2022,Elford2024}. HCN, HCO$^{+}$ and CS are instead the most used tracers of the dense molecular gas component, as their critical density (i.e.\,the density at which collisional excitation balances spontaneous radiative de-excitation) are $4.7\times10^5\,{\rm cm}^{-3}$, $6.8\times10^4\,{\rm cm}^{-3}$ and $1.5\times10^4\,{\rm cm}^{-3}$ respectively for their (1-0) transitions \citep[e.g.][]{Santi2014}. \cite{GaoSolomon2004a,GaoSolomon2004b} studied the HCN(1-0) line in a large sample of nearby normal star-forming (spiral)  and starburst galaxies and found a tight correlation between their infrared (IR; tracing star formation) and HCN luminosities. This has been interpreted as implying that the dense molecular gas is directly associated with active star forming regions, rather than being simply a tracer for the total molecular gas. Similar correlations have been later found by \cite{Tan2018}, but using both the HCN(4-3) and the $\rm HCO^+$(4-3) transitions as dense molecular gas tracers. 

Dense gas tracers have been detected also in AGN host galaxies \citep[e.g.][]{Baan2008,Krips2008,Juneau2009,Santi2014,Moser2016,Ruffa18,Impellizzeri19,Imanishi20,Li2021,Ruffa22}, and their ratios often used as a tool to probe the relative contribution of star formation and AGN to the excitation of the ISM.
For instance, a high intensity of HCN(1-0) with respect to $\rm HCO^+$(1-0) and/or CO(1-0) has been proposed as a feature unique to AGN. This is because the X-ray dissociation regions (XDRs) around radiatively-efficient AGN penetrate deep into the surrounding medium and destroy HCO+ molecules more efficiently than the analogous photodissociation regions (PDRs) in starburst areas \citep[e.g.][]{Jackson1993,Sternberg1994,Tacconi1994,Kohno2001,Usero2004,Kohno2005,Imanishi2007,Krips2008,Davies2012,Ruffa18}. Furthermore, \cite{Izumi2013} found that the HCN(4-3)/$\rm HCO^+$(4-3) and HCN(4-3)/CS(7-6) integrated intensity ratios are higher in AGN than in starburst galaxies. They thus proposed a diagnostic diagram based on these line ratios, which has been dubbed the "\textit{submm-HCN diagram}" (then expanded in \citealp{Izumi2016}). However, more work needs to be done to test how results from the \textit{submm-HCN diagram} compare to other methods of studying excitation/ionisation from AGN (e.g.\,by using the optical BPT diagrams; \citealt{Baldwin1981}) and fully understand the connection between nuclear activity and dense gas tracers.
\par
The Close AGN Reference Survey (CARS; \citealt{Husemann2017,Husemann2019}) is using facilities such as the Multi Unit Spectroscopic Explorer (MUSE; \citealp{Bacon2010,Bacon2014}) and ALMA to study 41 of the most luminous type 1 (i.e.\,unobscured) AGN at redshifts $0.01<z<0.06$.
In this paper, we present new ALMA observations of the CO, HCN, $\rm HCO^+$ and CS molecular gas transitions in 5 AGN of the CARS sample, which were selected to span the full range of AGN luminosity ($\log L_{\rm AGN}=42.9 - 45.4~\rm erg\,s^{-1}$) and stellar masses ($\rm 8.85<log\,M_*<11.2 \, M_\odot$) of the sample. We study the position of these sources on the Gao and Solomon relations \citep{GaoSolomon2004b,GaoSolomon2004a} between the IR and HCN luminosities. We also make use of the \textit{submm-HCN diagram} to examine its prediction for the cold gas excitation mechanism in these 5 objects. We then compare their classification in the sub-mm to the one in the optical using the classic BPT diagnostic diagram, that we create from the MUSE data available for the CARS sample.  
\par
The paper is organised as follows: in Section \ref{Observations}, we describe the sample and the observations used for our analysis. We describe the methodology and results in Sections \ref{Methods} and \ref{Results} respectively. We discuss our results in Section \ref{Discussion}, before summarising and concluding in Section \ref{Conclusion}.

\section{Observations}
\label{Observations}
In this study we use both interferometric data from ALMA and integral field unit (IFU) data from MUSE, all obtained as part of CARS. Full details about this survey can be found in \cite{Husemann2017}. In short, the CARS targets are drawn from the Hamburg-ESO Survey (HES, \citealt{Wisotzki2000}), which is a purely flux-limited ($B_J\lesssim17.3$) catalog of 415 luminous type 1 (unobscured) AGN based on \textit{B}-band optical photometry and slitless spectroscopy. Applying a redshift cut of $z\leq 0.06$, the HES catalog leads to a sample of 99 AGN. From these, a representative sub-sample of 41 objects was randomly selected and observed first in CO(1-0) with the IRAM-30m telescope \citep{Bertram2007}, and then with MUSE \citep{Husemann2017}. 
\par
The CARS sample is ideal for this type of study as they have spatially resolved observations and unobscured observations of the warm ionized medium (WIM) in nearby luminous type 1 AGN.
\par
\subsection{ALMA} 
The 5 CARS targets of this study are HE0108-4743, HE0433-1028, HE1029-1831, HE1108-2813 and HE1353-1917 (see Table \ref{tab:Gal_prop} for a summary of their main properties) which were chosen to span the full range of galaxy types and properties.\textbf{ The 2-10 keV X-ray luminosities were taken from \cite{Ricci2017} where available. However, for HE0108-4743 and HE1029-1831 we scale the available 0.1-2.4 keV X-ray observations to the 2-10 keV band assuming a spectral index of -0.8 (the mean spectral index reported by \citealp{ReevesTurner00}).} ALMA band 7 observations of the dense molecular gas component in these objects were taken in 2018, targeting the HCN(4-3), $\rm HCO^+$(4-3) and CS(7-6) lines (PI: Davis, Project Code: 2017.1.00258.S). \par
The spectral configuration of the ALMA observations consisted of 4 spectral windows (SPWs), three centred on the redshifted frequencies of the three targeted lines (354.505~GHz, 356.734~GHz and 342.883~GHz for HCN(4-3), HCO+(4-3) and CS(7-6), respectively) and one used to map the continuum. The integration times for these observations are 1391, 393, 1754, 3084 and 2933s for HE0108-4743, HE0433-1028, HE1029-1831, HE1108-2813 and HE1353-1917 respectively. For three of the five targets (i.e.\,HE0433-1028, HE1108-2813 and HE1353-1917), previous ALMA band 3 observations of the $\rm \rm ^{12}$CO(1-0) and $\rm ^{13}$CO(1-0) lines are also available (PI: Tremblay, Project Code: 2016.1.00952.S). The spectral configuration of the CO observations consisted of 4 SPWs: 2 centred on the redshifted frequency of the two targeted lines (115.271~GHz and 110.201~GHz for $^{12}$CO(1-0) and $\rm ^{13}$CO(1-0), respectively), and the other two used to observe the continuum.
For both sets of data a standard calibration strategy was adopted: a single bright quasar was used as both flux and bandpass calibrator, a second one as a phase calibrator. All data sets were calibrated using the the Common Astronomy Software Applications (CASA) pipeline, version 5.1.1 \citep{McMullin2007}. The obtained calibrated datasets were then combined (where needed) and imaged using the same CASA version.
\begin{figure*}
    \centering
    \includegraphics[width=0.9\textwidth]{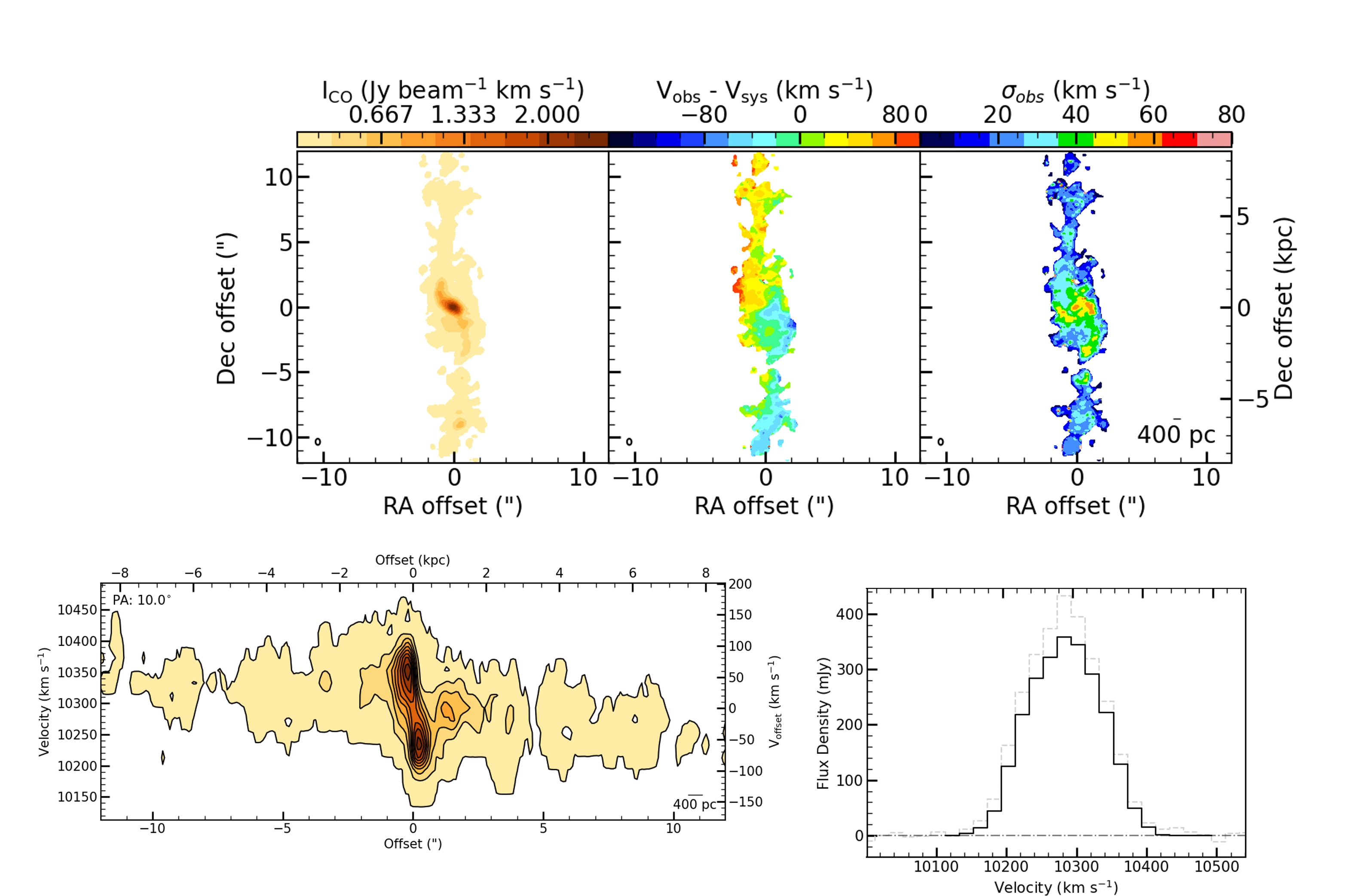}
    \caption{Data products of the $^{12}$CO(1-0) ALMA observations of HE0433-1028. The moment 0 (integrated intensity), moment 1 (mean line-of-sight velocity) and moment 2 (mean line-of-sight velocity dispersion) maps are shown in the top-left, top-middle and top-right panels, respectively. The position-velocity diagram (PVD) is shown in the bottom-left and the integrated spectrum in the bottom-right panels. The synthesised beam is shown in the bottom-left corner of each moment map, and a scale bar in shown in the bottom-right corner of the moment 2 map and the PVD. The bar above each moment map illustrates the colour scale.
    The integrated spectrum is extracted within a box of $2\arcsec \times 24\arcsec$. The line profile obtained from the masked data cube is illustrated by the black line, while the dashed line shows the line profile within the same region but without the mask being applied. The position angle (PA) was measured counterclockwise from the North direction. Coordinates are with respect to the image phase centre; East is to the left and North to the top.} 
    \label{fig:HE0433CO}
\end{figure*}

\subsubsection{Line Imaging}
Continuum emission was measured over the full line-free bandwidth and subtracted from the data in the \textit{uv}-plane using the {\sc CASA} task {\tt uvcontsub}. The line data was then imaged using the \textsc{casa} task {\tt tclean} with Briggs weighting and a robust parameter of 0.5, which allows to obtain the best trade-off between sensitivity and resolution. The resulting 3D (RA, Dec, velocity) data cubes were produced with channel widths of $\rm 20\,km\,s^{-1}$ and a pixel size which approximately Nyquist samples the synthesized beam. The HCN(4-3) and HCO$^{+}$(4-3) lines were clearly detected in three out of five sources (i.e.\,HE0433-1028, HE1029-1831 and HE1108-2813). $^{12}$CO(1-0) was also clearly detected in all the sources in which it was targeted (i.e.\,HE0433-1028, HE1108-2813 and HE1353-1917). CS(7-6) and $\rm ^{13}$CO(1-0) were not detected in any of the targets. The achieved angular resolution and 1$\sigma$ root mean square (rms) noise levels (measured in line-free channels) for each target and each molecular line are listed in Table \ref{tab:obs_param}. 
\subsubsection{Continuum imaging}
For each target, the continuum SPWs and the line-free channels of the line SPWs were used to produce the continuum maps using the {\sc CASA} task {\tt tclean} in multi-frequency synthesis (mfs) mode, and Briggs weighting with robust parameter of 0.5. This resulted in synthesised beam sizes in the range $0\farcs704-0\farcs863$ (corresponding to physical scales of $373-633$~pc) for the band 7 continuum maps, and $0\farcs447-0\farcs587$ ($284-426$~pc) for those in band 3. The systemic velocities ($\rm V_{sys}$) for HE0108-4743, HE0433-1028, HE1029-1831, HE1108-2813 and HE1353-1917 are 7171$\,{\rm km\,s}^{-1}$, 10294\,${\rm km\,s^{-1}}$, 11745\,${\rm km\,s}^{-1}$, 7027\,${\rm km\,s^{-1}}$ and 10103\,${\rm km\,s}^{-1}$ respectively and are estimated using the flux weighted velocity average from the data cubes. This average differs from the redshift listed in Table \ref{tab:Gal_prop} by less then 5\% in all our sources. The corresponding 1$\sigma$ rms noise levels are in the ranges $45.2-560$ $\rm \mu Jy \ beam^{-1}$ (band 7) and $11.4-20.8$ $\rm \mu Jy \ beam^{-1}$ (band 3). A single unresolved, point-like continuum source is detected at all frequencies and for almost all the targets. The only exception is the band 7 continuum map of HE0433-1028, where two point-like sources are observed: one coincident with the galaxy nucleus and the second one lying $\sim4\arcsec$ ($\sim3$~kpc) to the South of it. An overlay between the CO and band 7 continuum detections of this object allows us to verify that this second continuum source is also well within the CO disk (as can be seen in the moment maps in Figure~\ref{fig:HE0433CO}). However, due to the quite large separation of the two continuum detections and the second one not being detected in any other observation of this object, we conclude that this latter is likely a background galaxy. All continuum fluxes and associated uncertainties are reported in Table \ref{tab:cont_param}.\\
\subsection{MUSE}\label{sec:MUSE_data}
IFU data of the five targets was obtained with the MUSE instrument at the Very Large Telescope (VLT; \citealp{Bacon2010,Bacon2014}) under programs 094.B-0345(A) (HE0108-4743, HE0433-1028, HE1029-1831 and HE1108-2812) and 095.B-0015(A) (HE1353-1917). Here we use the extracted fluxes of the $\rm H\alpha$, $\rm [OIII] \lambda 5007$, $\rm H\beta$ and $\rm [NII] \lambda 6583$ lines.  Full details on the MUSE observations and data reduction can be found in \cite{Husemann2021}. In summary the MUSE observations had spatial resolutions ranging from $0\farcs43-1\farcs32$ and a pixel size of $0\farcs2$. In this work, we make use of the reduced MUSE data products, which are also publicly-available in the CARS archive\footnote{\url{https://cars.aip.de/}}.\\

\protect\begin{table*}
    \centering
    \caption{Basic parameters of our galaxy sample.}
    \begin{tabular}{cccccccccccccc}
         \hline
         Galaxy & z & ${\rm log}L_{\rm AGN}$ & ${\rm log}L_{\rm 2-10\,keV}$ & ${\rm log}M_{\star}$ & SFR & PA\\
         & & ($\rm erg\,s^{-1}$) & ($\rm erg\,s^{-1}$) & ($M_{\odot}$) & ($\rm M_{\odot} \, yr^{-1}$) & $^\circ$\\
         (1) & (2) & (3) & (4) & (5) & (6) & (7)\\
         \hline
         HE0108-4743 & 0.024 & 43.6 & 43.0 & 9.77 & $4.3^{+0.2}_{-0.2}$ & 90 \\ [0.1cm]
         HE0433-1028 & 0.036 & 44.8 & 43.3 & 10.80 & $19.4^{+0.2}_{-0.2}$ & 10.0\\ [0.1cm]
         HE1029-1831 & 0.041 & 44.3 & 43.4 & 10.49 & $27.1^{+0.8}_{-1.3}$ & 90.0\\ [0.1cm]
         HE1108-2813 & 0.024 & 44.0 & 42.7 & 10.29 & $12.4^{+0.4}_{-0.4}$ & 7.0\\ [0.1cm]
         HE1353-1917 & 0.035 & 44.1 & 43.4 & 10.99 & $2.8^{+0.1}_{-0.1}$ & 29.1 \\
         \hline
    \end{tabular}
    \parbox[t]{\textwidth}{\textit{Notes:} (1) galaxy name and its redshift based on the stellar continuum in the observed IFU data in (2). (3) AGN luminosity estimated from the H$\beta$ luminosity, (4) the 2-10\,keV X-ray luminosity of the object with the stellar mass of the object inferred from SED modeling in (5). (6) star formation rate with associated errors taken from \cite{Smirnova-Pinchukova2022}. (7) is the position angle of the object estimated from the CO moment 1 map for HE0433-1831 and HE1108-1813, taken from \cite{Husemann2019} for HE1353-1917 or taken from the NASA Extragalactic Database (NED)\textsuperscript{2} measured counterclockwise from North.\\
    \small\textsuperscript{2}\url{https://ned.ipac.caltech.edu/}}.
    \label{tab:Gal_prop}
\end{table*}

\begin{figure*}
    \centering
    \includegraphics[width=0.9\textwidth]{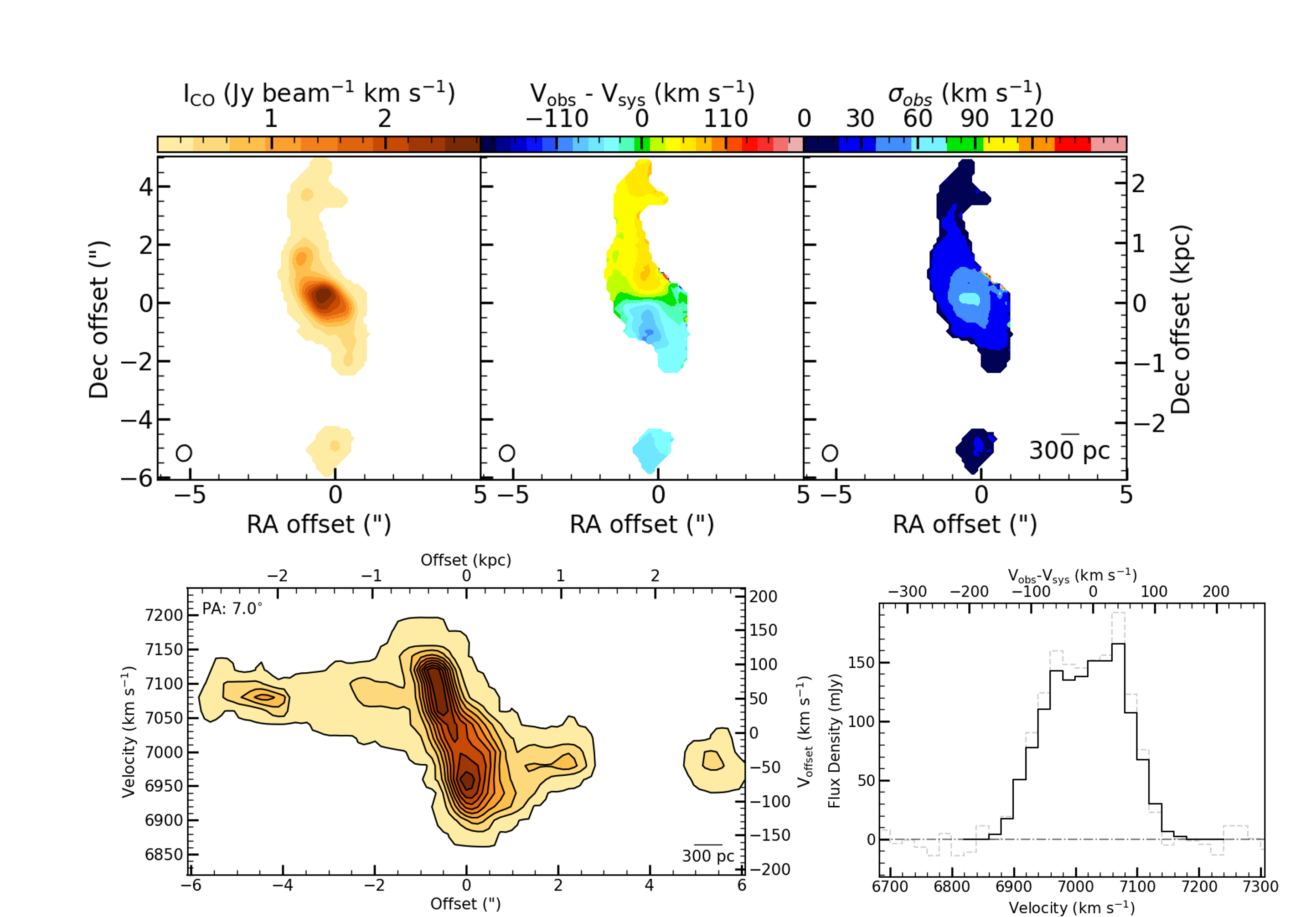}
    \caption{As in Figure \ref{fig:HE0433CO}, but for HE1108-2813. The spectrum is extracted within a box of $5\farcs0 \times 13\farcs5$.}
    \label{fig:HE1108CO}
\end{figure*}
  
\begin{table*}
\centering
\caption{Main properties of the presented ALMA observations.}\label{tab:obs_param}
\resizebox{\textwidth}{!}{
    \begin{tabular}{ccccccccccc}
        \hline
         Galaxy & $\rm \theta_{HCN}$ & $\rm \theta_{HCO+}$ & $\rm \theta_{CS}$ & $\rm \theta_{CO}$ & $\rm \theta_{^{13}CO}$ & $\sigma_{\rm HCN}$ & $\sigma_{\rm HCO+}$ & $\sigma_{\rm CS}$ & $\sigma_{\rm CO}$ & $\sigma_{\rm ^{13}CO}$  \\
         & (arcsec) & (arcsec) & (arcsec) & (arcsec) & (arcsec) & ($\rm mJy \ beam^{-1}$) & ($\rm mJy \ beam^{-1}$) & ($\rm mJy \ beam^{-1}$) & ($\rm mJy \ beam^{-1}$) & ($\rm mJy \ beam^{-1}$) \\
         (1) & (2) & (3) & (4) & (5) & (6) & (7) & (8) & (9) & (10) & (11)\\
         \hline
         HE0108-4743 & 0.731 & 0.727 & 0.759 & - & - &  0.624 & 0.686 & 0.646 & - & -\\
         HE0433-1028 & 0.899 & 0.900 & 0.936 & 0.406 & 0.417 & 1.26 & 1.29 & 1.43 & 0.492 & 0.455 \\
         HE1029-1831 & 0.735 & 0.735 & 0.763 & - & - & 0.673 & 1.38 & 1.00 & - & - \\
         HE1108-2813 & 0.789 & 0.789 & 0.818 & 0.520 & 0.528 & 0.831 & 0.883 & 0.837 & 0.895 & 0.744 \\
         HE1353-1917 & 0.694 & 0.692 & 0.719 & 0.533 & 0.557 & 0.466 & 0.499 & 0.547 & 0.374 & 0.351 \\
         \hline
         
    \end{tabular}}
    \parbox[t]{\linewidth}{\textit{Notes:} (1) galaxy name. (2) - (6) the geometric mean of the major and minor axis of the beam for HCN, HCO$^{+}$, CS, CO and $\rm ^{13}CO$ observations and corresponding $1\sigma$ rms noise levels in (7)-(11).}
\end{table*}

\begin{table}
\centering
\caption{Main properties of the ALMA continuum detections.}\label{tab:cont_param}
 \begin{tabular}{ccccc}
        \hline
         Galaxy & $S_{\rm cont, 339}$ & $\Delta S_{\rm cont, 339}$ & $S_{\rm cont, 104}$ & $\Delta S_{\rm cont, 104}$\\
         & (mJy) & (mJy) & (mJy) & (mJy)\\
         (1) & (2) & (3) & (4) & (5)\\
         \hline
         HE0108-4743 & 1.05 & 0.0534 & - & - \\
         \multicolumn{1}{p{2cm}}{\centering HE0433-1028\\(Main source)} & 2.25 & 0.122 & 0.727 & 0.0146 \\
         \multicolumn{1}{p{2cm}}{\centering HE0433-1029\\ (Secondary source)} & 1.89 & 0.122 & - & - \\
         HE1029-1831 & 5.56 & 0.0700 & - & - \\
         HE1108-2813 & 5.46 & 0.560 & 0.117 & 0.0208\\
         HE1353-1917 & 0.318 & 0.0452 & 0.185 & 0.0114 \\
         \hline
         
    \end{tabular}
    \parbox[t]{\linewidth}{\textit{Notes:} (1) galaxy name. (2) - (3) integrated flux densities of the band 7 continuum maps (reference frequency: 339~GHz) and corresponding uncertainties. (4) - (5) integrated flux densities of the band 3 continuum maps (reference frequency: 104~GHz) and corresponding uncertainties.}\end{table}

\begin{figure*}
    \centering
    \includegraphics[width=0.3\textwidth]{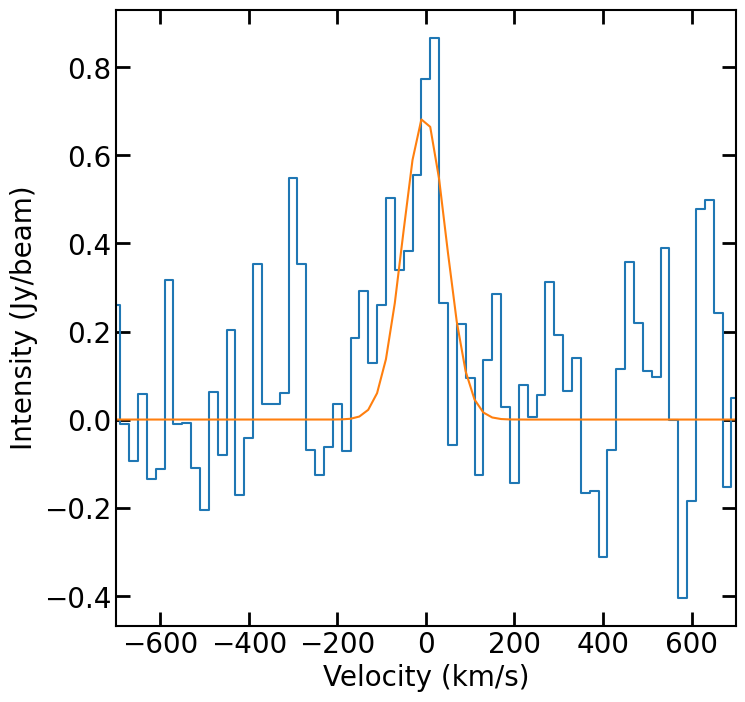}
    \includegraphics[width=0.3\textwidth]{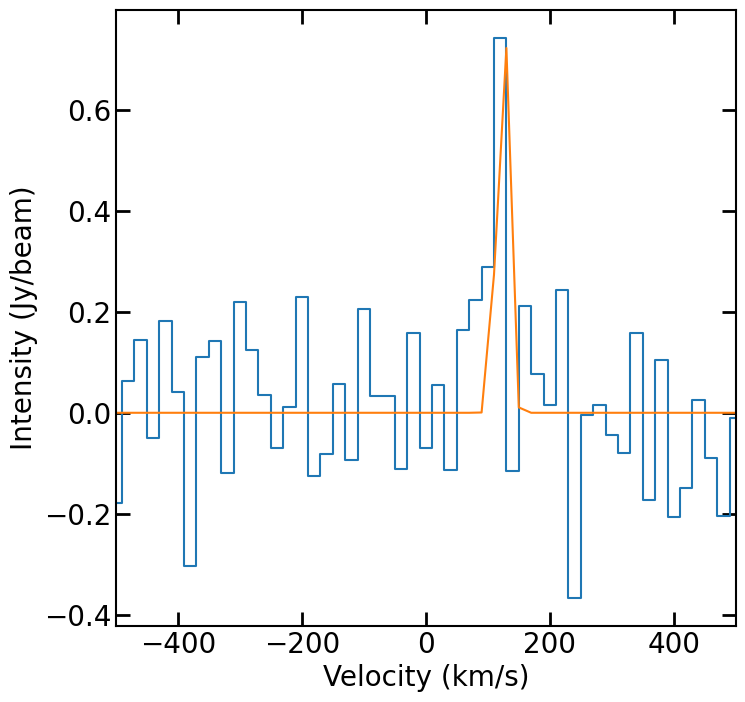}
    \includegraphics[width=0.3\textwidth]{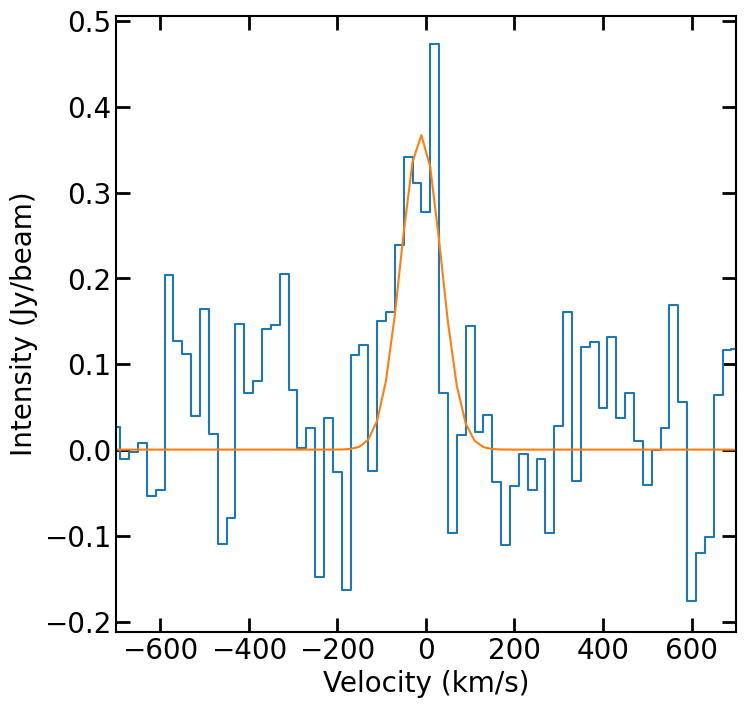}
    \caption{Stacked spectra of the HCN (4-3) line in HE0433-1028 (left), HE1029-1831 (middle) and HE1108-2813 (right). The orange lines shows the best-fitting Gaussian profiles.}
    \label{fig:HCN_Stack}
\end{figure*}

\begin{figure*}
    \centering
    \includegraphics[width=0.3\textwidth]{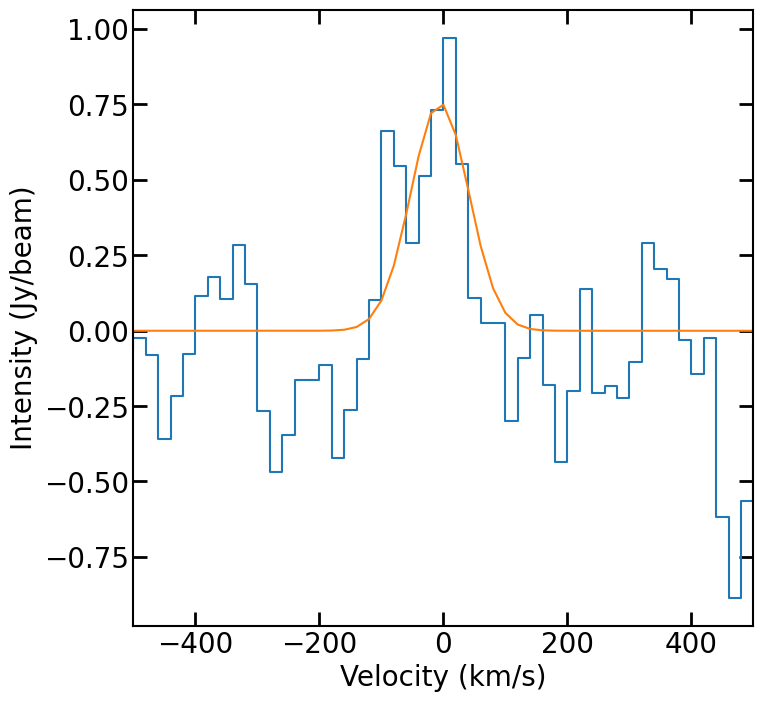}
    \includegraphics[width=0.3\textwidth]{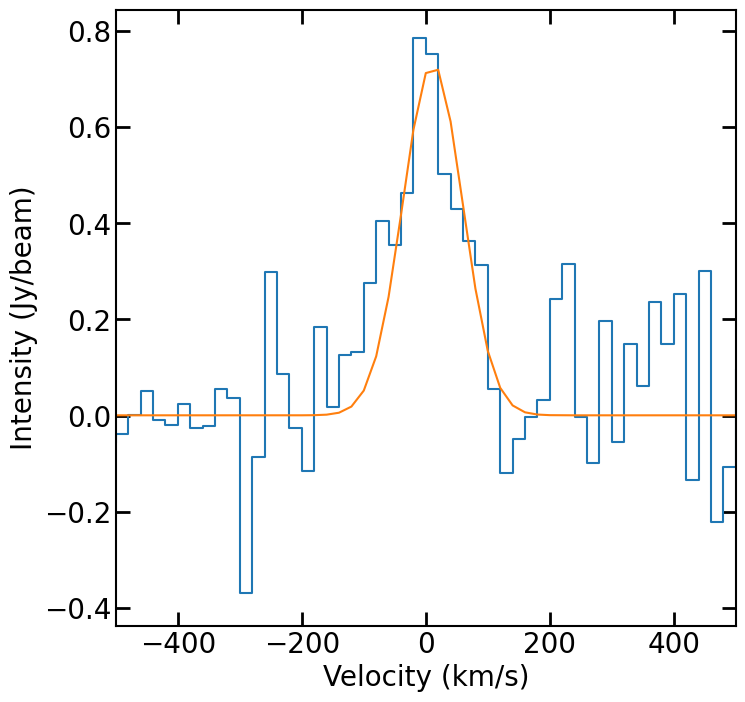}
    \includegraphics[width=0.3\textwidth]{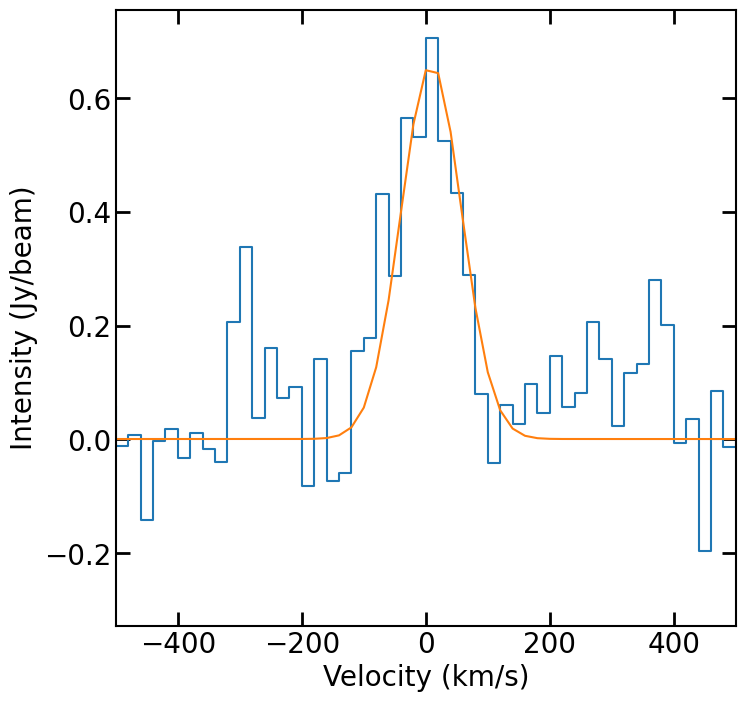}
    \caption{Stacked spectra of the HCO+ (4-3) line in HE0433-1028 (left), HE1029-1831 (middle) and HE1108-2813 (right). The orange lines shows the best-fitting Gaussian profiles.}
    \label{fig:HCO+_Stack}
\end{figure*}

\begin{figure*}
    \centering
    \includegraphics[width=0.3\textwidth]{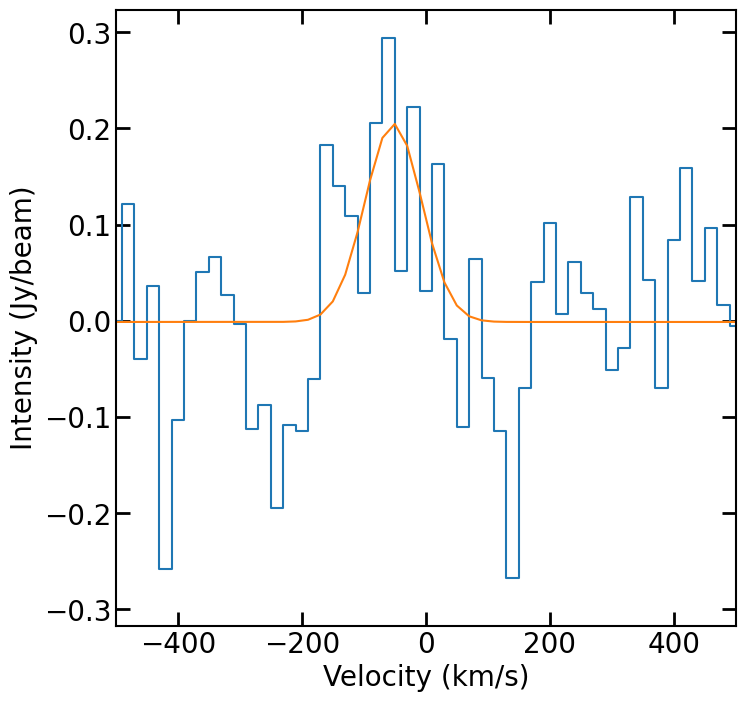}
    \caption{Stacked spectra of the CS (7-6) line in HE1108-2813. The orange line shows the best-fitting Gaussian profile.}
    \label{fig:CS_Stack}
\end{figure*}

\begin{figure*}
    \centering
    \includegraphics[width=0.45\textwidth]{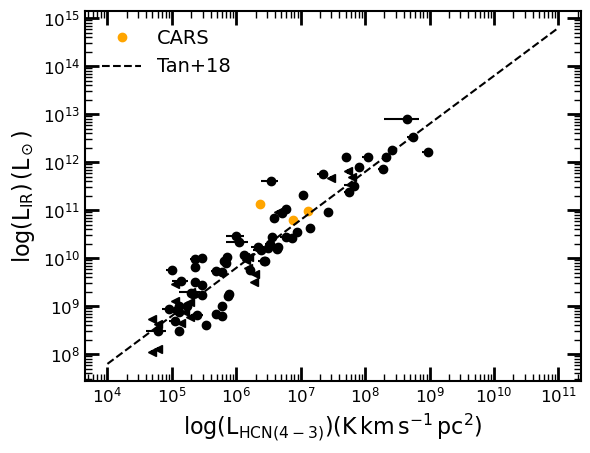}
    \includegraphics[width=0.45\textwidth]{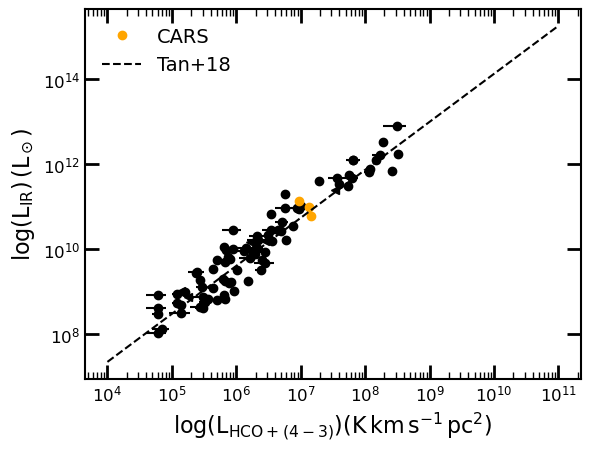}
    \caption{Relation between the HCN(4-3) (left) and HCO$^+$(4-3) (right) luminosity and the IR luminosity from star formation. We additionally plot the points and relation from \protect \cite{Tan2018}.}
    \label{fig:GS_relation}
\end{figure*}

\begin{figure}
    \centering
    \includegraphics[width=0.5\textwidth]{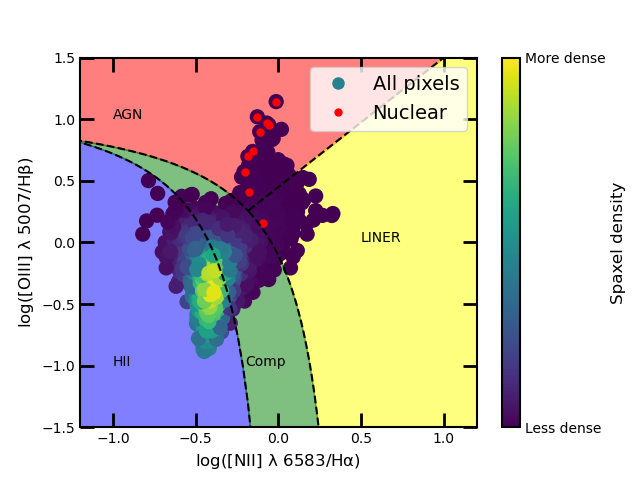}
    \includegraphics[width=0.5\textwidth]{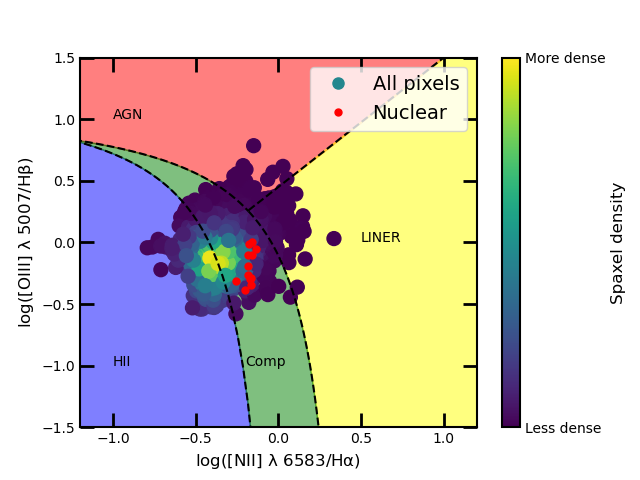}
    \includegraphics[width=0.5\textwidth]{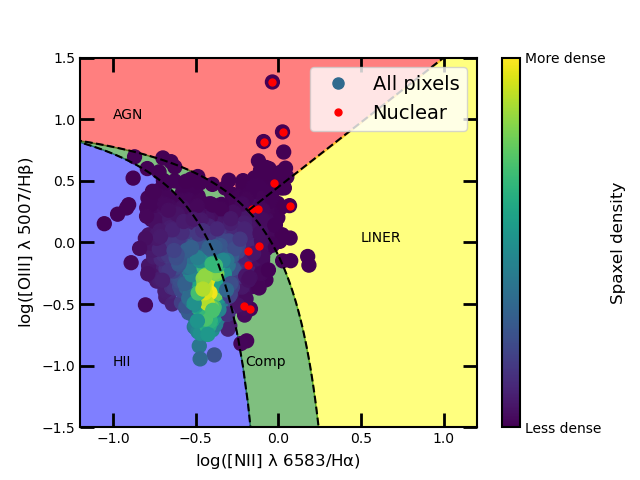}
    \caption{All pixels BPT diagrams of HE0433-1028 (top), HE1029-1831 (middle) and HE1108-2813 (bottom). The lighter regions are areas with a higher density of points. The red points show the results obtained for the nuclear regions of each object, assumed to be within one ALMA beam size. The blue, green, yellow and red regions are where the star formation, composite, LINERs and AGN dominate the excitation respectively.}
    \label{fig:BPT}
\end{figure}

\section{Methods}
\label{Methods}
\subsection{ALMA data products}\label{ALMA Data}
Integrated intensity (moment 0), mean line-of-sight velocity (moment 1) and velocity dispersion (moment 2) maps of the HCN, HCO$^{+}$, CS and CO detections were produced from the continuum-subtracted data cubes using the masked moment technique \citep{Dame2011}, as implemented in the {\sc Pymakeplots} routine\footnote{\url{https://github.com/TimothyADavis/pymakeplots}}. In this procedure, a copy of the cleaned data cube is first Gaussian-smoothed spatially with a full width at half maximum (FWHM) equal to 1.5 times that of the synthesised beam, and then smoothed in the velocity dimension with a boxcar kernel FWHM of 4 channels. A three-dimensional mask is then created by selecting all the pixels above a fixed flux-density threshold (chosen by the user) in the smoothed data cube. The moment maps are then constructed using the original, un-smoothed cubes within the masked regions only. In the majority of the cases, we adopted a flux-density threshold for the mask of $5\sigma$, to ensure that only real, high-significance gas emission is included. Major-axis position velocity diagrams (PVDs) of the line detections were then constructed by summing the pixels in the masked cube within a pseudo-slit whose long axis is oriented along the position angle of gas distribution. 
Integrated spectra were created from the observed data cubes by summing over both spatial dimensions within boxes enclosing all the detected line emission. The obtained data products are illustrated in Figures~\ref{fig:HE0433CO} and \ref{fig:HE1108CO} for the CO(1-0) detections in  HE0433-1028 and HE1108-2813, respectively (the data products for HE1353-1917 have already been presented in \citealp{Husemann2019} and thus not reproduced here), and in Figures~\ref{fig:HE0433HCN}-\ref{fig:HE1108HCO+} for all the HCN(4-3) and HCO$^+$(4-3) detections.

The CO(1-0) emission in HE0433-1028 and HE1108-2813 is well resolved. In particular, the integrated intensity maps (top-left panels of Figures \ref{fig:HE0433CO} and \ref{fig:HE1108CO}) shows that in both cases the CO gas is distributed along the bar of the galaxy and extends $\approx 22'' (\sim16.4 \,{\rm kpc}) $ and $\approx 13.5'' (\sim6.7 {\rm kpc})$, respectively, along its major axis. The integrated flux densities of the two CO(1-0) detections are listed in Table \ref{tab:ALMA_data_stack}. These were estimated by numerically integrating over the channels defining the corresponding line profiles (illustrated in the bottom-right panels of Figures \ref{fig:HE0433CO} and \ref{fig:HE1108CO}). The intensity-weighted mean line-of-sight velocity maps (top-middle panels of Figures \ref{fig:HE0433CO} and \ref{fig:HE1108CO}) and intensity-weighted line-of-sight velocity dispersion maps (top-right panels of Figures \ref{fig:HE0433CO} and \ref{fig:HE1108CO}) suggest signs of shocks, such as regions of high velocity dispersion \citep[e.g][]{Athanassoula92}. This would occur as the molecular gas flows along the x1 orbits (i.e. elongated parallel to the bar) of the bar in both sources, and also along the x2 orbit (i.e. perpendicular to the bar) in HE1108-2813. This can be inferred as regions of high mean velocity and velocity dispersion are clearly visible in both cases along the leading edge of the bar and in the central regions, where the transition from x1 to x2 orbits occurs. The major-axis PVDs (bottom-left panels of Figures \ref{fig:HE0433CO} and \ref{fig:HE1108CO}) show that the CO gas disk is regularly rotating in both of these sources. There is also an asymmetric increase in the velocity in both of these sources which may be associated with non-circular motions in the elliptical x2 orbits at the centre of each system.

As visible in Figures~\ref{fig:HE0433HCN}-\ref{fig:HE1108HCO+}, the HCN(4-3) and HCO$^{+}$(4-3) emission is mostly unresolved in all the three targets in which these lines have been detected, and thus their moment maps do not yield much information. However, from these moment maps it is clear that the dense gas is concentrated in the nuclear regions ($\sim$197-336pc in radius, corresponding to half the beam width of the ALMA observations) of the galaxies where it co-exists with the less dense molecular gas.

\subsection{Stacking methodology} \label{Stack}
To attempt the detection of the weaker lines (and improve the signal-to-noise of those already detected), we performed a stacking analysis using the {\sc Stackarator} tool \footnote{\url{https://github.com/TimothyADavis/stackarator}} of Davis et al.\,(submitted). This routine allows to stack the datacubes of extended sources to extract weak lines by means of the velocity field (i.e.\,moment 1 map) of a bright line detection of the same source. For our targets, we used the velocity information of the $^{12}$CO(1-0) when available, the MUSE stellar velocity maps otherwise. Each spaxel in the cube was velocity-shifted according to the corresponding value in the reference moment 1 map, aligning the local spectra to their respective systemic velocities before stacking. The adjusted spectra of each pixel are then co-added with appropriate rebinning to create a single, stacked spectrum. This procedure allows us to improve the average signal-to-noise ratio (S/N) of the HCN(4-3) and HCO$^{+}$(4-3) detections in HE0433-1028, HE1029-1831 and HE1108-2813 from $\sim3.7$ to $\sim4.1$ (see Figures \ref{fig:HCN_Stack}--\ref{fig:CS_Stack}), whereas the other two targets remain undetected in these lines. The stacking of the CS datacubes allows us to pull out a faint detection in HE1108-2813 (as visible in Figure~\ref{fig:CS_Stack}). All of the other targets remain undetected in CS(7-6). $\rm ^{13}CO$ is not detected in any of the stacked datacubes. This is not surprising, as the ALMA band 3 observations were tailored to detect $\rm ^{12}CO$ (which is typically $\gtrsim10-20$ times brighter than its isotopologue). The integrated flux densities and FWHM of the detected HCN(4-3), $\rm HCO^+$(4-3) and CS(7-6) lines were obtained by performing a single Gaussian fit on their stacked spectra using the Markov chain Monte Carlo (MCMC) code {\sc GAStimator}\footnote{\url{https://github.com/TimothyADavis/GAStimator}}. The upper-limits for non-detections were estimated using the relation from \cite{Koay16}:
\begin{equation}
    I\le3\sigma \Delta v_{\rm FWHM}\sqrt{\frac{\Delta v}{\Delta v_{\rm FWHM}}},
\end{equation} where $\sigma$ is the rms noise level, $\Delta v$ is the channel width and $\Delta v_{\rm FWHM}$ is the expected FWHM of the emission line. In Table \ref{tab:ALMA_data_stack} we list the obtained values for all the targets.

\begin{table*}
    \caption{Main properties of the detected lines}
    \resizebox{\textwidth}{!}{
    \begin{tabular}{cccccccccccc}
         \hline
         Galaxy & $I_{\rm HCN (4-3)}$ & $FWHM_{\rm HCN (4-3)}$ & $I_{\rm HCO^+(4-3)}$ & $FWHM_{\rm HCO^+(4-3)}$ & $I_{\rm CS(7-6)}$ & $FWHM_{\rm CS(7-6)}$ & $I_{\rm CO (1-0)}$ & $I_{\rm CO^{13} (1-0)}$ & $\rm V_{shift,CO}$ & $\rm V_{shift,HCN}$ & $\rm V_{shift,HCO^+}$\\
         & $\rm (Jy\,km\,s^{-1})$ & $\rm (km\,s^{-1})$ & $\rm (Jy\,km\,s^{-1})$ & $\rm (km\,s^{-1})$ & $\rm (Jy\,km\,s^{-1})$ & $\rm (km\,s^{-1})$ & $\rm (Jy\,km\,s^{-1})$ & $\rm (Jy\,km\,s^{-1})$ & $(\rm km\,s^{-1})$ & $\rm (km\,s^{-1})$ & $\rm (km\,s^{-1})$\\
         (1) & (2) & (3) & (4) & (5) & (6) & (7) & (8) & (9)\\
         \hline
         HE0108-4743 & <0.93 & - & <1.09 & - & <1.17 & - & - & - & - & - & -\\
         HE0433-1028 & $2.25\pm0.11$ & 114 & $2.39\pm0.12$ & 111 & <1.58 & - & $56.5\pm0.03$ & <1.96 & 492 & -162 & -98\\
         HE1029-1831 & $0.60\pm0.09$& 20.5 & $2.48\pm0.16 $ & 113 & <1.23 & - & - & - & - & 111 & -171\\
         HE1108-2813 & $0.78\pm0.06$ & 106 & $1.53\pm0.06$ & 115 & $0.42\pm0.06$ & 108 & $25.2\pm0.05$ & <2.99 & 877 & 247 & -66\\
         HE1353-1917 & <0.38 & - & <0.50 & - & <0.51 & - & - & <2.05 & - & - & -\\
         \hline
    \end{tabular}}
    \label{tab:ALMA_data_stack}
    \parbox[t]{\textwidth}{\textit{Notes:} (1) Galaxy name. (2)-(6) HCN(4-3), HCO$^{+}$(4-3), CS(7-6), CO(1-0) and $\rm ^{13}CO$(1-0) line intensities and associated uncertainties. All but CO(1-0) values are those obtained after stacking (see Section~\ref{Stack}). (7) the velocity shift in the CO(1-0) line} 
    
\end{table*}
\section{Results}\label{Results}
\subsection{Testing the Gao-Solomon relation for AGN}\label{GS_Relation}

To better understand how the AGN can impact the dense gas surrounding it and how this in turn can impact star formation, we compare the dense gas luminosities with known star formation relations. \par
In Figure \ref{fig:GS_relation} we relate the HCN(4-3) and HCO$^{+}$(4-3) luminosities of HE0433-1028,  HE1029-1831 and HE1108-2813 with the infrared luminosities derived from their star formation rates.  We estimated the infrared luminosities of our targets by inverting the relation \citep{GaoSolomon2004a,Kennicutt98}
\begin{equation}
    {\rm SFR}\approx2\times10^{-10}(L_{\rm IR}/{\rm L_\odot}){\rm M_{\odot}}yr^{-1}.
\end{equation}.
The star formation rates are taken from \cite{Smirnova-Pinchukova2022} and are listed in Table \ref{tab:Gal_prop}.
We then used the revised version of the Gao and Solomon relation from \cite{Tan2018}:
\begin{equation}
    {\rm log}L_{\rm IR}=1.00(\pm0.04){\rm log}L'_{\rm HCN(4-3)}+3.80(\pm0.27)
\end{equation}
and
\begin{equation}
    {\rm log}L_{\rm IR}=1.13(\pm0.04){\rm log}L'_{\rm HCO^+(4-3)}+2.83(\pm0.24)
\end{equation}
which was calibrated on starburst galaxies, LIRGs and ULIRGs. Our sources have $L_{\rm IR}\sim10^{11}\,{\rm L_\odot}$ which would classify them as LIRGs. To confirm whether our sources follow the Gao and Solomon relations we performed a one-sample t-test on the residuals between the observed $L_{\rm IR}$ and that predicted from the Gao and Solomon relation for both HCN and HCO$^+$. We obtained p-values of 0.32 and 0.51 respectively indicating no significant deviations. Despite the potential impact of the AGN on these line-fluxes, we find that all the three CARS targets are consistent with following the Gao and Solomon relations. These strong relations suggest that star formation occurs in dense molecular gas. This will be discussed further in Section~\ref{Discussion}.\\

\subsection{Gas excitation classification}
\subsubsection{Optical BPT diagrams}\label{BPT_analysis}
To understand the dominant ionisation mechanism(s) of the warm ionised medium (WIM) in the 3 targets with corresponding ALMA data, we produced BPT diagrams using the available MUSE IFU CARS data (see Section~\ref{sec:MUSE_data} for details).\par
In a classic BPT diagram \citep{Baldwin1981}, the $\rm [NII]\lambda 6583/H\alpha$ and $\rm [OIII]\lambda 5007/H\beta$ ratios are used as proxies to classify objects as either star formation (HII), composites, low-ionisation nuclear emission region (LINER) or AGN dominated. In particular, to discriminate between star formation-dominated and composite regions, we used the following equation (derived from the observations presented in \citealp{Kauffmann2003}):
\begin{equation}
    \rm log([OIII]/H\beta=0.61/(log([NII]/H\alpha)-0.05)+1.3.
\end{equation}
On the other hand, to separate between composite and AGN/LINER regions, we adopted the following relation (defined from photoionisation and stellar population synthesis modeling in \citealp{Kewley2001}):
\begin{equation}
    \rm log([OIII]/H\beta=0.61/(log([NII]/H\alpha)-0.47)+1.19.
\end{equation}
We then used the equation presented in \cite{Schawinski2007} to distinguish between AGN and LINER regions:
\begin{equation}
    \rm log([OIII]/H\beta=1.05log([NII]/H\alpha)+0.45.
\end{equation}
\par

All pixels BPT diagrams of the targets were produced using the definitions illustrated above. In addition, we also produced the same diagnostic diagrams for the nuclear regions, which were obtained by including only the area of the IFU data cubes within one ALMA beam. This was done to probe the scales that are most likely to be impacted by the AGN, ensuring also a comparison on similar scales both in the optical and the sub-mm. The obtained all pixels and nuclear BPT diagrams are shown in Figure \ref{fig:BPT}, where each point represents one pixel in the MUSE maps. The colours of the all pixel BPTs indicate the density of points in an area, with brighter points for denser areas.

The BPT diagnosis identifies the AGN as the dominant ionisation mechanism in the nuclear regions of HE0433-1028, and a combination of star formation and AGN (i.e.\,composite) in those of HE1029-1831 and HE1108-2813. This thus suggests that, in these latter sources, the optical excitation has large contributions from star formation alongside the impact of the bright AGN. This is not especially surprising, as these objects are highly star forming, with SFRs of 10-30 M$_{\odot}$ yr$^{-1}$ (see Table \ref{tab:Gal_prop}).

Additionally, we studied the AGN diagnostic diagrams proposed in \cite{Veilleux1987}, the so-called VO87-OI and VO87-SII. These use the $\rm [OI]\lambda 6300/H\alpha$, $\rm [OIII]\lambda 5007/H\beta$ or $\rm [SII]\lambda67177/H\alpha$ ratios to classify objects as either star forming, LINER or Seyfert-dominated. We find that the VO87-OI and VO87-SII diagrams are in good agreement with the BPT classification illustrated in Figure \ref{fig:BPT}, thus we do not show or further discuss them here.

\subsubsection{sub-mm HCN diagram}\label{submm_classification}
\label{HCN_Diag}
\begin{figure*}
    \centering
    \includegraphics[width=0.9\textwidth]{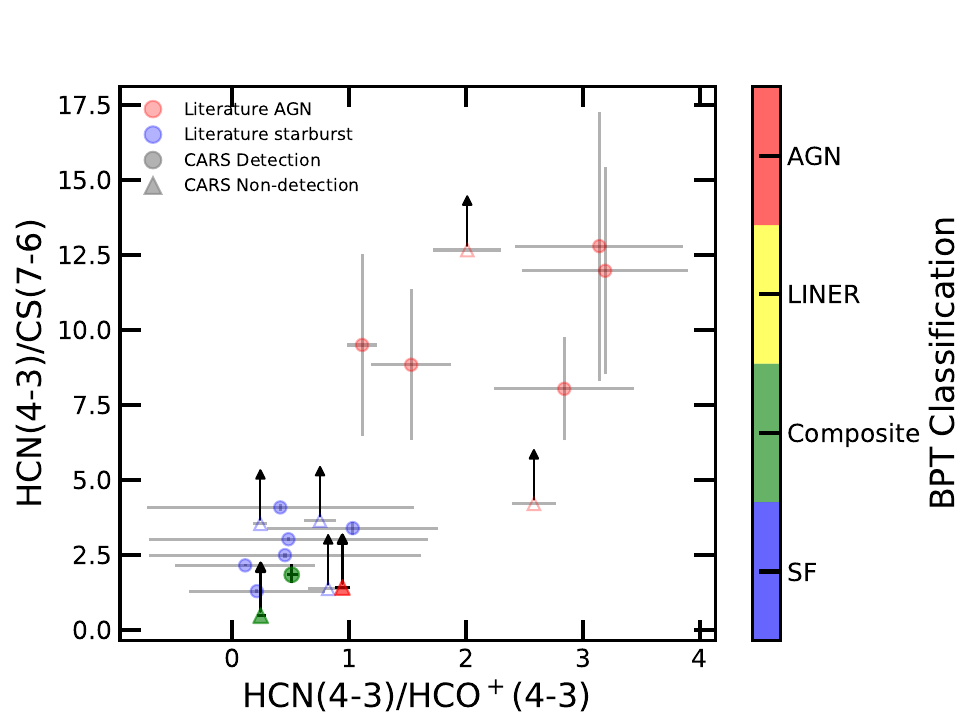}
    \caption{Submm-HCN diagnostic diagram. The three CARS targets for which this analysis has been carried out are plotted in green and red, as labelled by the colour bar on the right (and based on their BPT classification). We also plot on the background the data from \protect \cite{Izumi2016}. Triangles represent lower limits with open triangle representing literature lower limits. This plot suggests that the Submm-HCN diagram and the optical BPT diagram are in agreement with each other. The CARS sources that are identified as AGN using the BPT diagram lie in the same region as the literature non-detected AGN and in reality likely have higher HCN(4-3)/CS(7-6) ratios taken moving them up into the AGN region of the diagram. On the other hand, the sources that are identified as composite using the BPT diagram (i.e. have significant contribution from star formation lie in same region as the starburst sources from literature.}
    \label{fig:submm_diag}
\end{figure*}
To make a comparison with the optical classification, we produce the submm-HCN diagnostic diagram using the HCN(4-3)/CS(7-6) and HCN(4-3)/HCO+(4-3) ratios. As described in Section~\ref{ALMA Data}, we detect all these three lines only in HE1108-2813, as CS(7-6) is not detected in HE0433-1028 and HE1029-1831. For these latter sources, we used upper limits on the CS(7-6) line to place lower limits on the HCN(4-3)/CS(7-6) ratio. 

The submm-HCN diagram with these three targets included is shown in Figure \ref{fig:submm_diag}, with the points coloured based on their BPT classification. We additionally plot the data from \cite{Izumi2016}. We find all three of our sources fall in the starburst region of the diagram though the two lower limits could move upwards. In HE1108-2813, the galaxy in which we detect all the required lines, the two diagrams agree on the excitation mechanism, as both predict star formation excitation to be important. Our other two systems only have lower limits on the HCN/CS axis, making it harder to judge agreement between the two diagrams. Both sources have lower HCN/HCO$^+$ ratios than any of the AGN sources from \cite{Izumi2016}. The source HE0433-1028 which is classified as AGN dominated in the optical BPT diagram is, however, the most similar to the AGN in this ratio. If its true HCN/CS line ratio were significantly higher than our lower limit, then within error it would be fully consistent with AGN excitation dominating also in the dense gas. These results will be further discussed in Section \ref{Discussion}.

\section{Discussion}
\label{Discussion}
The large scale molecular discs of the 5 AGN from CARS show a variety of different morphologies, with bars being important in HE0433-1028 and HE1108-2813. The denser gas tracers are only detected towards the nuclei, close to the AGN.\par 
In Section \ref{GS_Relation} we found that the 3 sources with HCN(4-3) and HCO$^+$(4-3) detections follow known star formation relations. This indicates that if the HCN abundance has been enhanced due to the presence of an AGN, as speculated by \cite{Izumi2016}, it has not been enhanced enough to effect global scaling relations in log-log space.

In Sections  \ref{BPT_analysis} and \ref{submm_classification} we compared the BPT and sub-mm diagnostics to identify the excitation mechanism of the optical ionised and molecular gas lines. We find all three of our sources fall in the starburst region of the sub-mm HCN diagram though the two lower limits could move upwards. For HE1108-2813, the galaxy in which we detect all the required lines, the two diagrams agree on the excitation mechanism, as both predict star formation excitation to be important. Our other two systems only have lower limits on the HCN/CS axis, making it harder to judge agreement between the two diagrams. If the true HCN/CS ratio of HE0433-1028 (our source which is classified as AGN dominated in the optical BPT diagram) were significantly higher (e.g. a ratio of $\gtrsim8$) than our lower limit, then within error it would be fully consistent with AGN excitation dominating also in the dense gas. This would then mean the optical BPT diagram and sub-mm HCN diagram are in good agreement with each other. We stress that this crosscheck has only been performed on a very small sample of three sources and this, along with the non-detection of some lines, makes it hard to draw solid general conclusions. Further comparisons with deeper observations on larger samples are needed to study this further.
\par
In optical classifications like the BPT the difference in ionised gas line ratios in star-forming and active galaxies is due to the conditions of the ISM, the gas-phase abundance and the different ionisation processes. In star-forming galaxies the line ratios are reproduced by photoionisation models that have O and B stars as the main sources of ionisation \citep{Kewley2001,Nagao06,Stasinska06,Levesque10,Sanchez15}. AGN typically have higher ratios because the accretion processes can produce higher-energy photons which can in turn induce more heating in the narrow line region (NLR), thus causing an increase in the collisionally excited lines \citep{Stasinska06}. This explains why our three sources, which are all type 1 AGN (selected via the presence of broad lines, not using these line ratios), show signs of AGN ionisation in their BPT diagrams (see Figure \ref{fig:BPT}).

By using non local thermodynamic equilibrium (LTE) radiative transfer models, \cite{Izumi2016} found that the high HCN to HCO$^+$ and HCN to CS dense molecular line ratios of AGN are likely due to their enhanced HCN abundance compared to star forming galaxies. This also indicates a higher gas density in AGN with respect to star forming galaxies. The exact mechanism that causes this HCN abundance in AGN is not clear with \cite{Izumi2016} proposing several potential explanations, including high-temperature chemistry ($\rm \sim300K$), the effects of metallicity and chemical abundances, IR pumping or time dependent chemistry as potential solutions.

Our findings are consistent with the idea that AGN can be an important excitation mechanism at sub-mm wavelength. It indicates that the AGN radiation can penetrate both the low density ionised gas (as probed by the BPT classification) and the high density molecular gas (as probed by the sub-mm HCN classification). It seems that the hard X-ray photons produced by nuclear activity can both ionise the diffuse gas and penetrate deep into the denser molecular gas clouds, heating them \citep[e.g.][]{Maloney96,Wolfire22}. This heating could then induce high temperature chemistry that leads to the increase in HCN abundance as discussed in \cite{Izumi2016}. 

For comparison, \cite{Esposito22} studied the CO emission in a sample of 35 local active galaxies to investigate the main source of molecular gas heating in these sources. In their study they found that on the scales of $\approx$250\,pc from the AGN a combination of star formation and AGN ionisation is needed to reproduce the CO emission observed, in agreement with what we find here.  

On the other hand, some works have shown that AGN radiation is absorbed by the outer layers of dense gas clouds which prevents the AGN radiation from penetrating deep into the molecular cloud \citep{Roos15,Bieri17,Meenakshi22}. If this remains true in our sources then perhaps the AGN could be heating the gas by inducing shocks in the clouds that can excite the dense gas \citep{Namekata14,Meenakshi22}. These shocks could be produced by either outflows or jets which have been identified in some CARS sources \citep{Husemann2019,Singha2023}. However, we caution that these shocks would need to be affecting the dense and ionised gas simultaneously, which seems less likely given that the dense and ionised gas may not be co-spatial.

Furthermore, the AGN could produce cosmic rays or could be driving turbulence potentially through outflows, which results in large line-width, and produce turbulence/velocity shears in the ISM \citep[e.g.][]{Singha2022}. Both of these processes have been shown to substantially impact molecular clouds \citep[e.g.][]{Federrath12,Federrath13,Gaspari18,Crocker21,Wittor23}. In particular, the density fluctuations imparted by AGN-driven turbulence are key to promote the next-generation of condensing gas via nonlinear thermal instability, leading to the chaotic cold accretion of such dense clouds that recursively stimulate the AGN feeding-feedback cycle (\citealt{Gaspari20}, for a review).

Finally, the fact that we are not seeing an enhanced HCN excitation could be due to the resolution of our observations. It has been found in \cite{Viti14} that the HCN excitation is enhanced in the circumnuclear disc $\rm r<200\,pc$ and \cite{Audibert21} was able to separate out the AGN HCN excitation thanks to the high resolution ($\sim$4pc) beam. Our ALMA beams which have sizes (373-633pc) may be too large to separate out the AGN HCN excitation with the AGN impact being diluted. This is further supported by AGN in the low-resolution sample from \cite{Izumi16}, which where their AGN sources fall in the starburst region of the sub-mm HCN diagram. Following on from this, it could be the case that the AGN has a higher weight in the BPT diagram, meaning its impact is not diluted like it is for the HCN excitation, explaining why we still see the AGN impact in the nuclear BPT diagrams.

Overall, our results are consistent with the idea that AGN may be able to directly or indirectly excite the gas in dense molecular clouds. However, when strong star formation is present this can easily dominate the excitation, of both gas phases, even close to a luminous AGN. 

\section{Conclusions}
\label{Conclusion}
Using CARS ALMA data we have observed dense and molecular gas tracers in 5 nearby type 1 (unobscured) AGN.
In summary:
\begin{itemize}
    \item We have found that the large scale molecular discs of these systems show a variety of different morphologies, with bars being important in several cases. The denser gas is only detected towards the nuclei, close to the AGN where the gas density is supposedly the highest. We have used the observations of the dense gas tracers to study their use as an AGN diagnostic tool.
    \item It has been reported in previous studies that the HCN/HCO+ and HCN/CS ratios can be used to separate AGN from star forming galaxies using a \textit{sub-mm HCN diagram}, as HCN can be enriched by chemical pathways at high temperatures. We find that all three of the sources where the dense gas tracers were detected follow the Gao-Solomon relations between the amount of dense gas and star-formation. This indicates that if the HCN abundance has been enriched by AGN activity it has not been enhanced enough to affect global scaling relation. 
    \item When studying the \textit{sub-mm HCN diagram} we are limited by the lack of detection of CS(7-6) in many of our sources, but in HE1108-2813 we find general agreement between optical and sub-mm classification gas excitation mechanisms. It should be noted that these results are limited due to the low statistics and the interpretation is speculative.
\end{itemize}

This suggests that AGN can contribute to the excitation of both the low density gas in the WIM and high density molecular gas clouds in the nuclear regions ($\sim$250pc in radius), perhaps through X-ray, cosmic ray or shock heating mechanisms. 

\par Further observations of dense gas around AGN are clearly needed to further populate the sub-mm HCN diagram and to allow further cross-comparisons with other AGN classification schemes and studies of AGN excitation at different wavelengths. Deeper observations with longer integration times are key to detecting CS (7-6) to be able to accurately place our sources and future sources on the sub-mm HCN diagram. Observations of other sub-mm lines could also help to understand the impact of AGN on the dense gas that surrounds it with \cite{Imanishi16} finding the HCN-to-HCO$^+$ J=3-2 flux ratio also being higher in AGN hosting galaxies.  

\section*{Acknowledgements}
This work is supported by the UKRI AIMLAC CDT, funded by grant EP/S023992/1. TAD and IR acknowledges support from STFC grant ST/S00033X/1. 
M.G. acknowledges support from the ERC Consolidator Grant \textit{BlackHoleWeather} (101086804). CO and SB acknowledge support from the Natural Sciences and Engineering Research Council (NSERC) of Canada. 
\par
This research made use of ASTROPY\footnote{\url{http://www.astropy.org/}},  a community-developed PYTHON package for Astronomy \citep{Astropy2013}, MATPLOTLIB\footnote{\url{https://matplotlib.org/}}, an open source visualisation package \citep{Matplotlib2007} and NUMPY\footnote{\url{https://numpy.org/}}, an open source numerical computation library \citep{Numpy2020}
\par
This paper makes use of ALMA data. ALMA is a partnership of
the ESO (representing its member states), NSF (USA), and NINS
(Japan), together with the NRC (Canada), NSC, ASIAA (Taiwan), and KASI (Republic of Korea), in cooperation with the Republic
of Chile. The Joint ALMA Observatory is operated by the ESO,
AUI/NRAO, and NAOJ.\par
This paper made up work towards the PhD thesis \cite{JacobElfordThesis}.

\section*{Data Availability}
The data underlying this article are available in
the ALMA archive, at \url{http://almascience.eso.org/aq/}. The reduced data used will be shared upon reasonable request to the first author.



\bibliographystyle{mnras}
\bibliography{example} 




\appendix

\section{ALMA observations of HCN(4-3) and HCO+(4-3)}
\begin{figure*}
    \centering
    \includegraphics[width=0.55\textwidth]{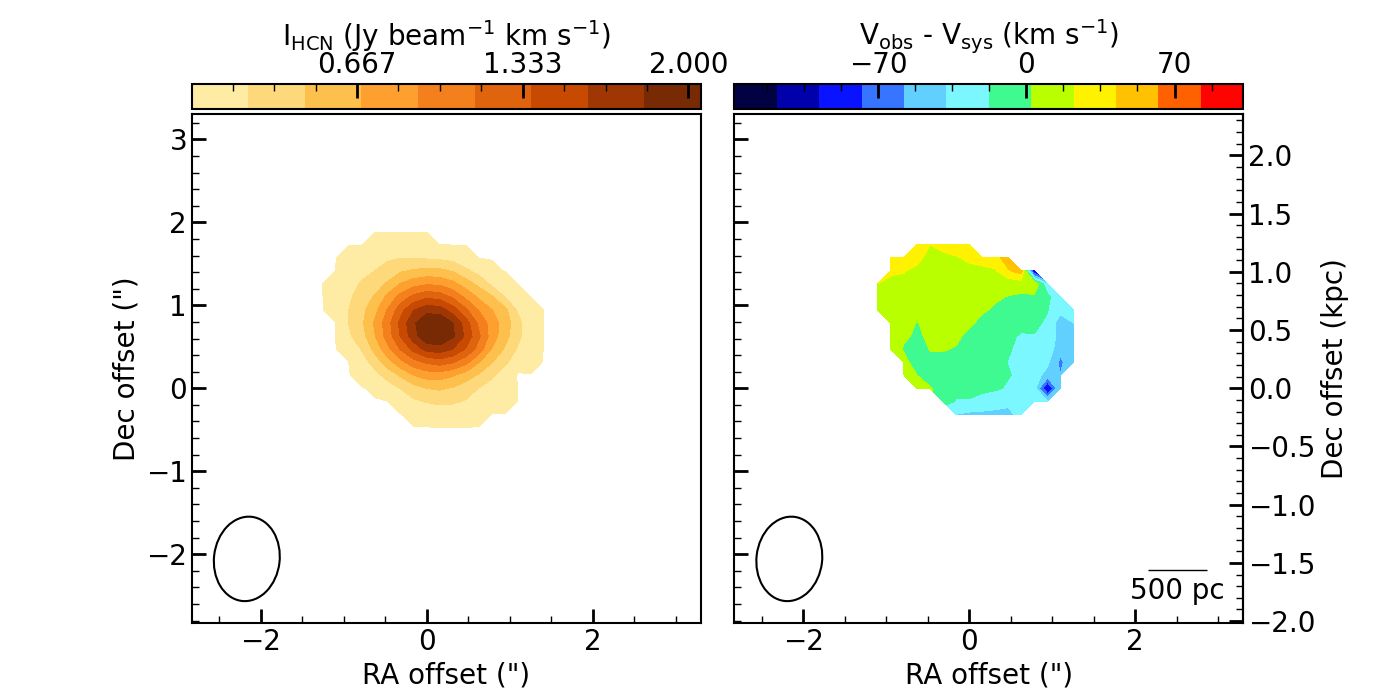}
    \includegraphics[width=0.35\textwidth]{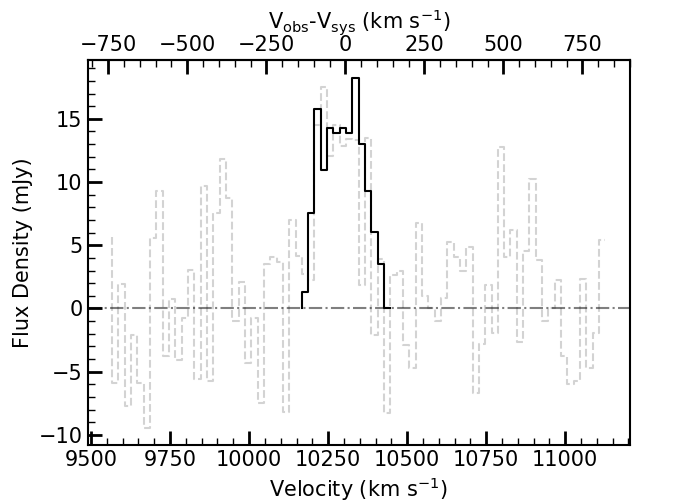}
    \caption{HCN moment 0 (left), moment 1 (middle) and integrated spectrum (right) obtained from the ALMA observations of HE0433-1028. The synthesised beam is shown in the bottom-left corners of each map and a scale bar is shown in the bottom-right corner of the moment 1 map. The bar above each map illustrates the colour scale. The spectrum has been extracted within a box of 3$\farcs$0$ \times 2\farcs5$. Coordinates are with respect to the image phase centre; East is to the left and North to the top. $1\farcs\approx$747pc.}
    \label{fig:HE0433HCN}
\end{figure*}

\begin{figure*}
    \centering
    \includegraphics[width=0.55\textwidth]{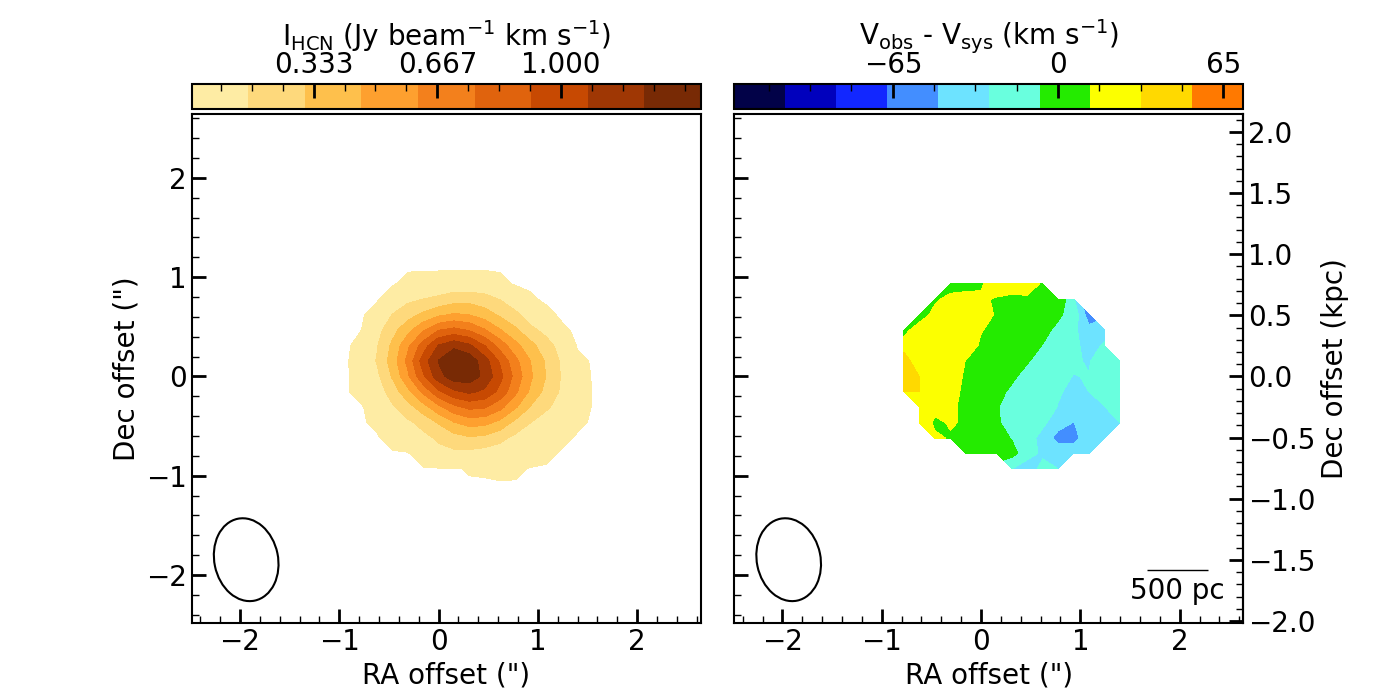}
    \includegraphics[width=0.35\textwidth]{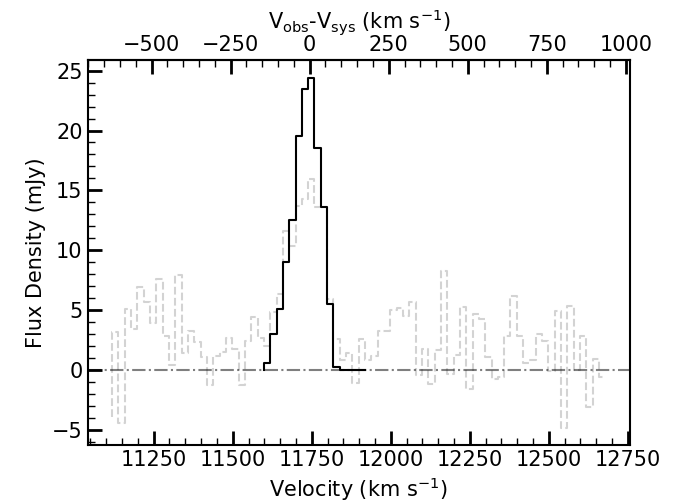}
    \caption{As in Figure~\ref{fig:HE0433HCN}, but for HE1029-1831. The spectrum has been extracted within a box of $2\farcs0 \times 1\farcs5$. $1\farcs\approx$850pc.}
    \label{fig:HE1029HCN}
\end{figure*}

\begin{figure*}
    \centering
    \includegraphics[width=0.55\textwidth]{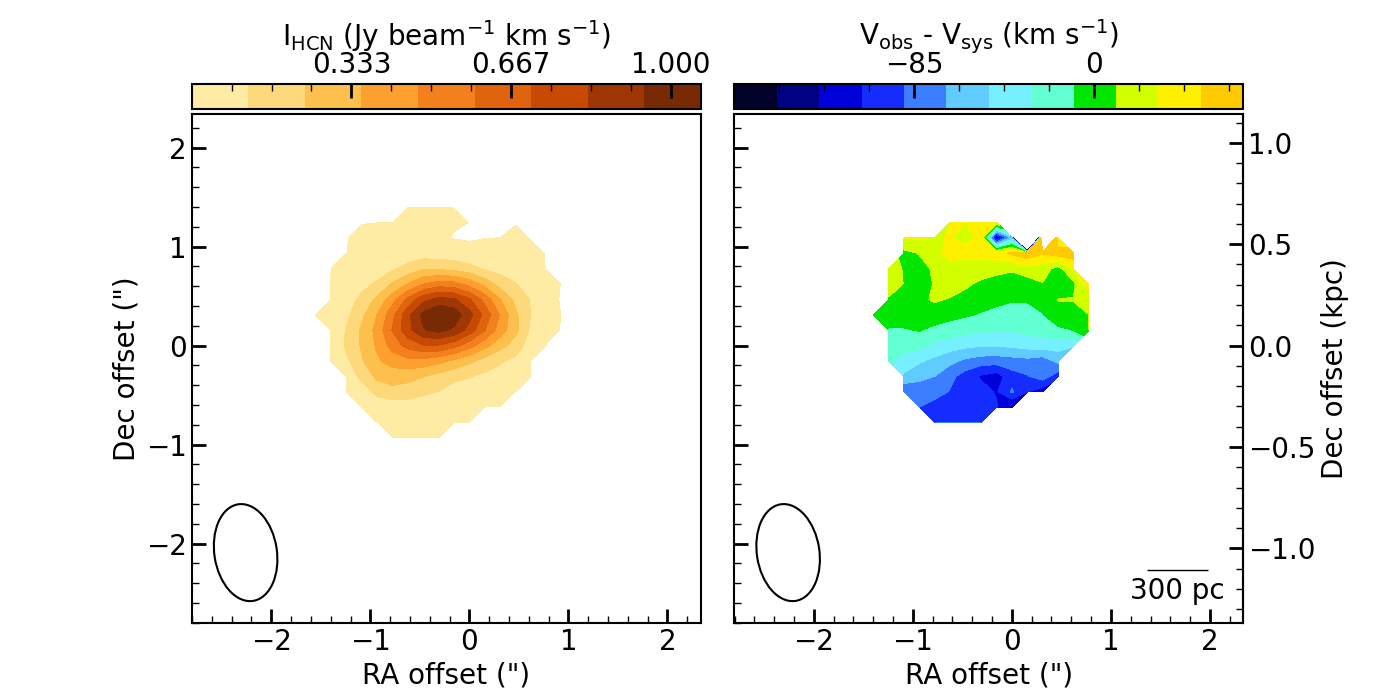}
    \includegraphics[width=0.35\textwidth]{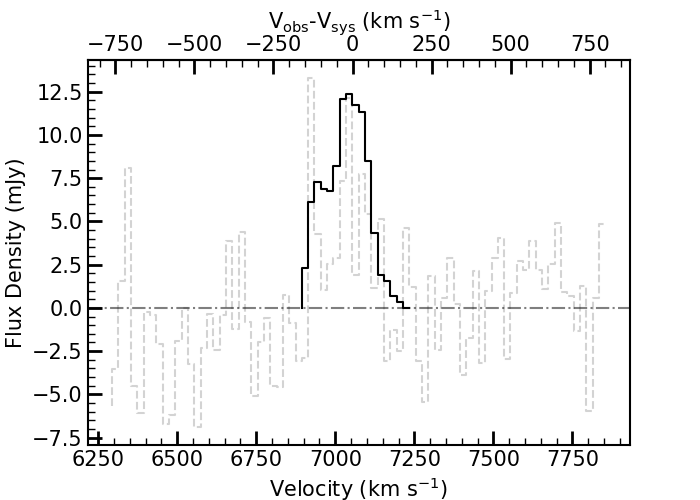}
    \caption{As in Figure~\ref{fig:HE0433HCN}, but for HE1108-2813. The spectrum  has been extracted within a box of $5\farcs25 \times 3\farcs75$. $1\farcs\approx$747pc.}
    \label{fig:HE1108HCN}
\end{figure*}

\begin{figure*}
    \centering
    \includegraphics[width=0.55\textwidth]{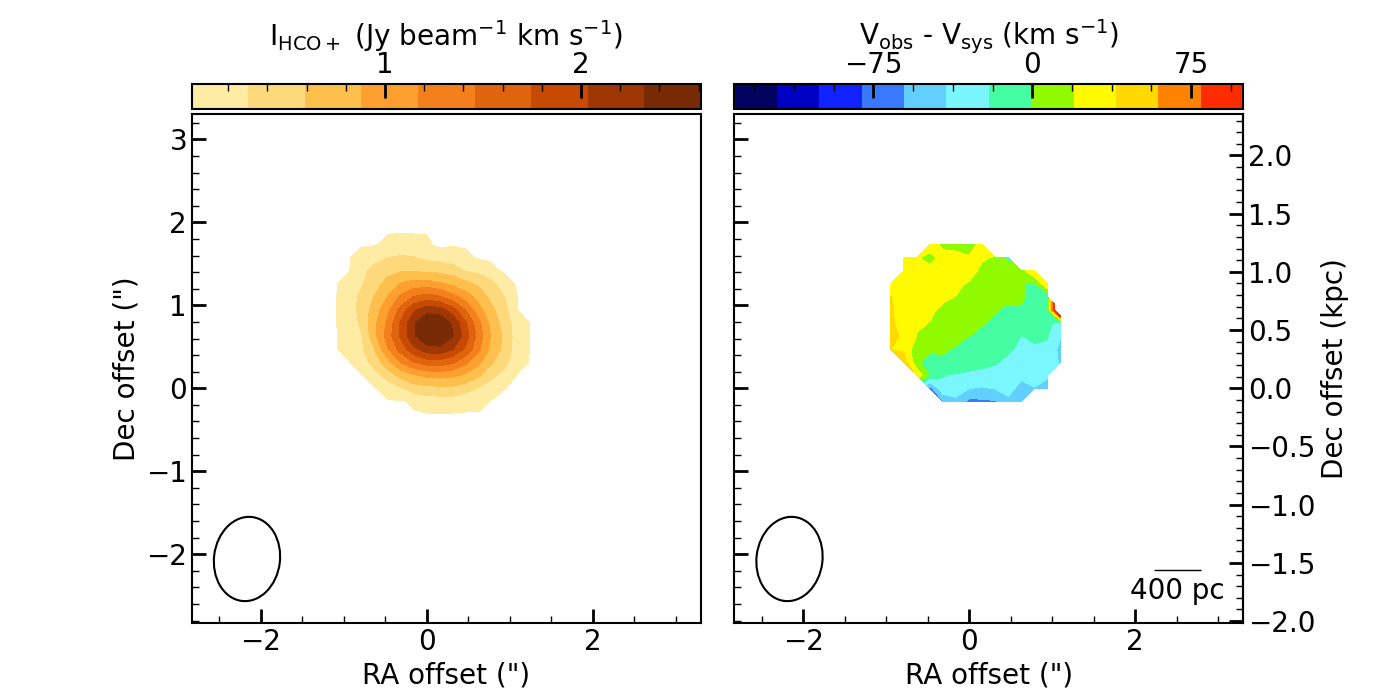}
    \includegraphics[width=0.35\textwidth]{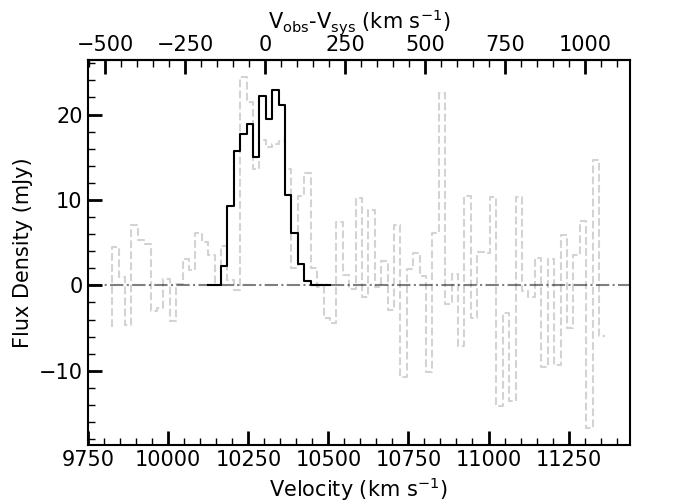}
    \caption{As in Figure~\ref{fig:HE0433HCN}, but for the HCO$^{+}$ detection of HE0433-1028. The spectrum has been extracted within a box of $1\farcs0 \times 2\farcs5$. $1\farcs\approx$747pc.}
    \label{fig:HE0433HCO+}
\end{figure*}

\begin{figure*}
    \centering
    \includegraphics[width=0.55\textwidth]{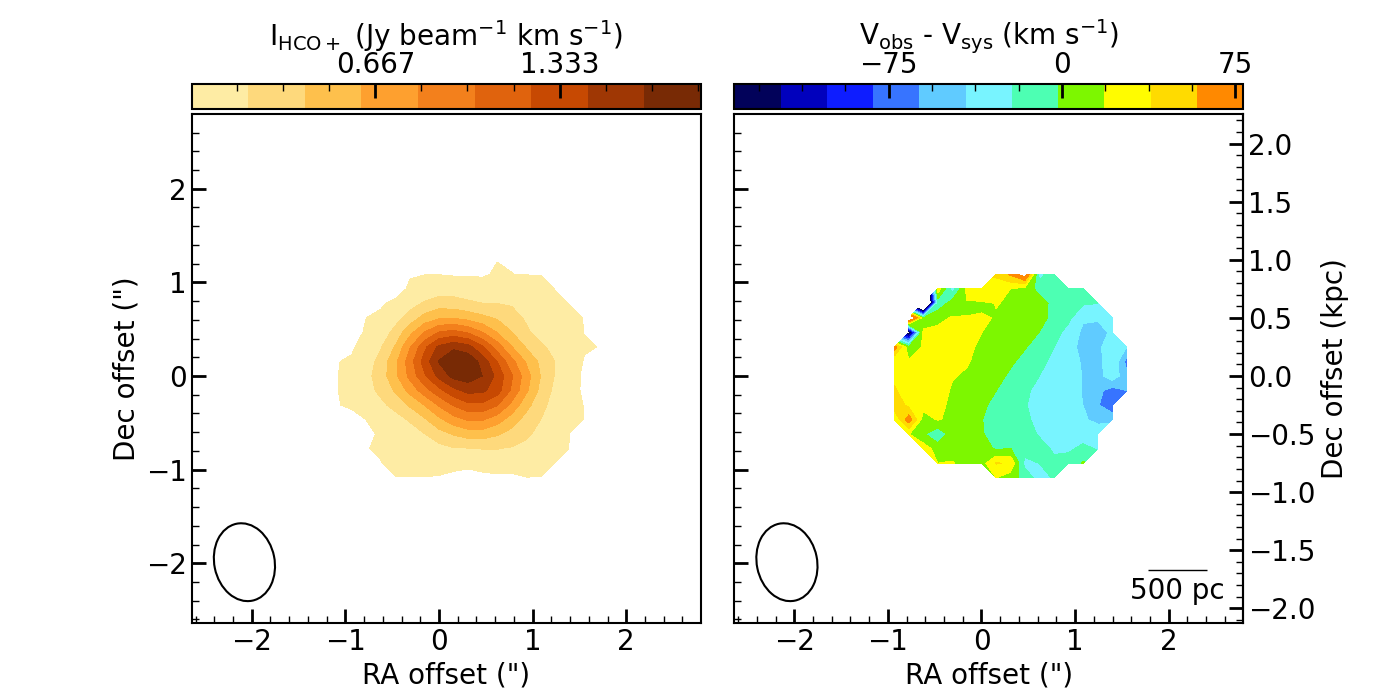}
    \includegraphics[width=0.35\textwidth]{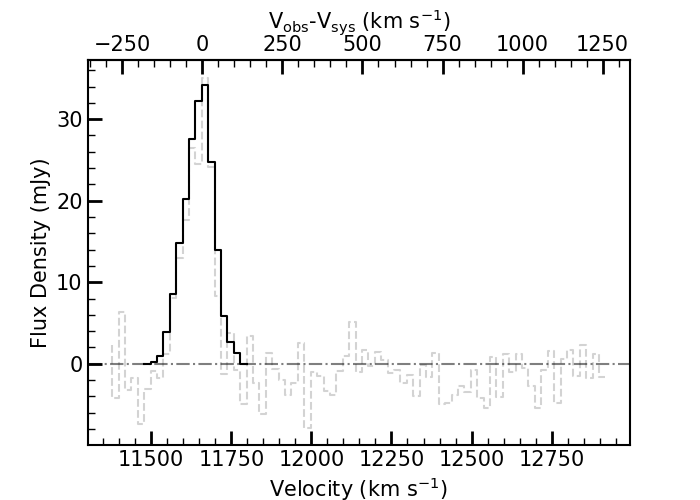}
    \caption{As in Figure~\ref{fig:HE0433HCN}, but for the HCO$^{+}$ detection of HE1029-1831. The spectrum has been extracted within a box of $3\farcs75 \times 2\farcs40$. $1\farcs\approx$850pc.}
    \label{fig:HE1029HCO+}
\end{figure*}

\begin{figure*}
    \centering
    \includegraphics[width=0.55\textwidth]{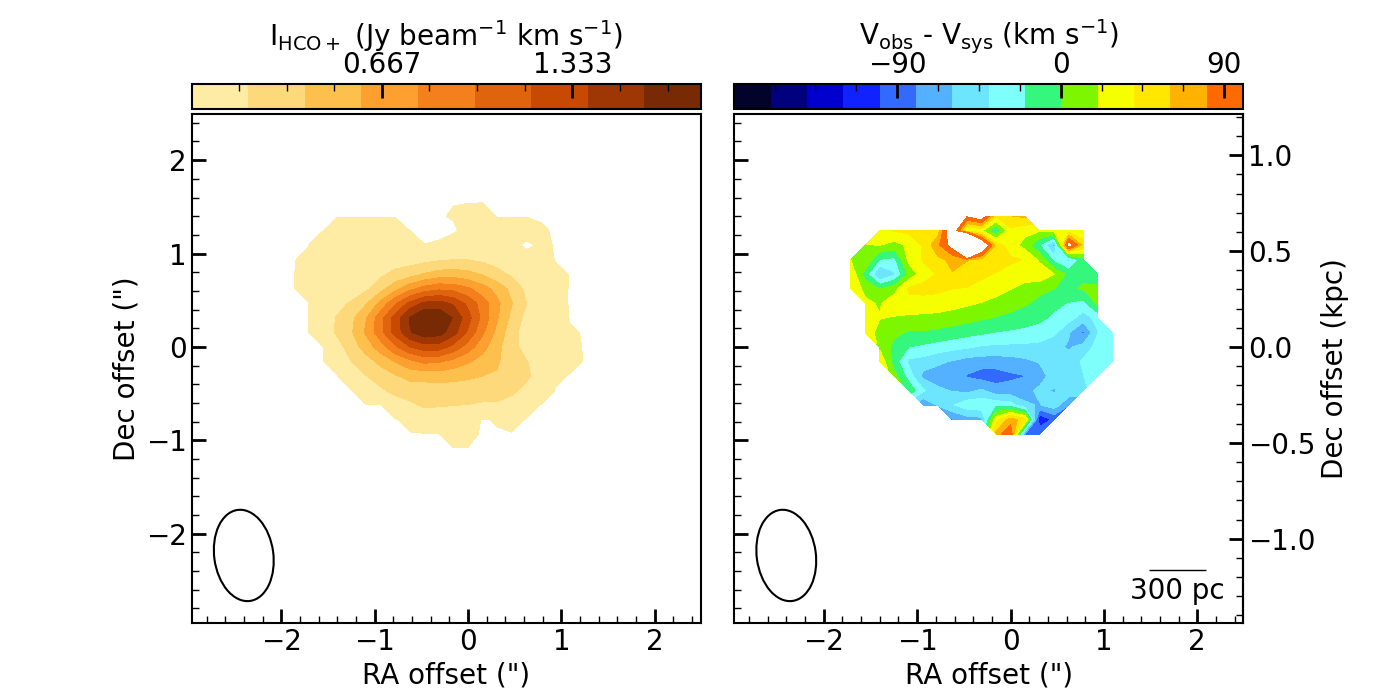}
    \includegraphics[width=0.35\textwidth]{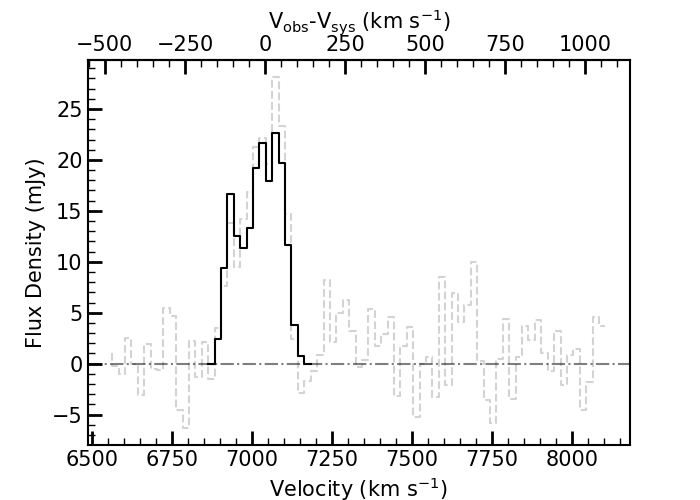}
    \caption{As in Figure~\ref{fig:HE0433HCN}, but for the HCO$^{+}$ detection of HE1108-2813. The spectrum has been extracted within a box of $3\farcs75 \times 3\farcs60$. $1\farcs\approx$747pc.}
    \label{fig:HE1108HCO+}
\end{figure*}

In this appendix we present the moment 0 and 1 maps along with the spectra for the ALMA observations of the HCN(4-3) (Figures \ref{fig:HE0433HCN}--\ref{fig:HE1108HCN}) and HCO$^+$ (Figures \ref{fig:HE0433HCO+}--\ref{fig:HE1108HCO+}) where it was detected in HE0433-1028, HE1029-1831 and HE1108-2813. In these three targets the emission is mostly unresolved and the moment maps do not yield much information. However, it is clear that the dense gas is concentrated in the nuclear regions.


\bsp	
\label{lastpage}
\end{document}